\documentclass[10pt]{article}
\usepackage[left=2.5cm, right=2.5cm, top=2.5cm, bottom=2.5cm]{geometry}

\usepackage{authblk} 
\usepackage{graphicx}
\usepackage{hyperref}
\usepackage{amsmath, amssymb, physics, nicefrac}
\usepackage[dvipsnames]{xcolor}
\usepackage{algorithm}
\usepackage{algpseudocode}
\usepackage{appendix}
\usepackage{afterpage}
\usepackage{titlesec}
\usepackage{lineno}
\usepackage{multirow}
\usepackage[normalem]{ulem}
\usepackage{tabularray}
\usepackage{makecell}

\usepackage[style=phys, sorting=none]{biblatex}
\newcommand{\keywords}[1]{%
  \par\vspace{0.5ex}%
  \noindent\textbf{Keywords: }#1\par
}

\title{A data-driven framework to identify restenosis-prone regions in femoral arteries from geometric and inflow waveform parameters}

\author[1,2]{Chotirawee Chatpattanasiri}
\author[2,3]{Federica Ninno}
\author[1,2]{Vanessa D\'{\i}az-Zuccarini}
\author[1,2,*]{Stavroula Balabani}

\affil[1]{\small \raggedright Department of Mechanical Engineering, University College London, London, UK}
\affil[2]{\small \raggedright Hawkes Institute, Department of Medical Physics and Biomedical Engineering, University College London, London, UK}
\affil[3]{\small \raggedright Department of Medical Physics and Biomedical Engineering, University College London, London, UK}
\affil[*]{\small \textit{Corresponding author:} \href{s.balabani@ucl.ac.uk}{s.balabani@ucl.ac.uk}}

\date{} 

\begin{document}

\maketitle

\begin{abstract}

Haemodynamic indices derived from Computational Fluid Dynamics (CFD), such as Time-averaged Wall Shear Stress (TAWSS) and Oscillatory Shear Index (OSI), are closely associated with restenosis risk in Peripheral Arterial Disease (PAD). However, translating these insights into clinical practice may require computationally efficient approaches such as Reduced Order Model (ROM) or Machine Learning (ML). To address this, we developed an ML-ROM framework to predict critical, restenosis-prone, haemodynamic regions accounting for both vessel geometries and inlet flow waveforms. We generated 500 synthetic femoral-artery geometries parameterised by six geometric parameters, and created physiologically realistic inflow waveforms via Principal Component Analysis (PCA) of patient data. CFD was used to obtain the Wall Shear Stress (WSS) distribution, from which TAWSS and OSI were computed. Critical regions were then defined by applying threshold-based criteria to the TAWSS and OSI. Four critical-region definitions were considered: two with vessel-specific relative thresholds (TAWSS$<33^{rd}$ percentile and OSI$>66^{nd}$ percentile) and two with absolute thresholds (TAWSS$<0.5$ Pa and OSI$>0.2$). Proper orthogonal decomposition (POD) was then applied to these high-dimensional critical-region data to obtain ROMs; These were then used to train ML models from which the critical region regions could be reconstructed. Three ML architectures were explored: a Fourier-based architecture, a Long Short-term Memory (LSTM) architecture, and a Convolutional Neural Network (CNN) architecture. The Fourier models achieved the highest performance, with the median values of Balanced Accuracy (BA) exceeding 0.92 across all critical-region definitions. The model also offered a substantial speed-up ratio in the order of $10^9$ compared to traditional CFD.

\end{abstract}

\keywords{Restenosis, Peripheral Artery Disease, Wall Shear Stress, Proper Orthogonal Decomposition, Reduced Order Model, Machine Learning.}

\section{Introduction}

Lower-limb Peripheral Arterial Disease (PAD) is a common vascular condition that affects over 200 million people worldwide, predominantly impacting the older population and individuals with cardiovascular risk factors \cite{fowkes2017peripheral, SHAMAKI2022101082, Houghton2024}. It is primarily caused by atherosclerosis, a progressive disease marked by plaque buildup that narrows and stiffens the arteries supplying the lower limbs. When symptoms become severe or limb-threatening, patients often undergo surgical or endovascular revascularisation procedures, such as bypass grafting, balloon angioplasty, or stent implantation \cite{SHAMAKI2022101082}. Despite initial procedural success, a substantial number of patients experience restenosis (re-narrowing) of the treated vessel, within months to years following intervention \cite{SHAMAKI2022101082, Ninno2024modelling_lower_limb, gokgol2019prediction, colombo2020computing, colombo2021baseline, colombo2022superficial, carbonaro2023design, corti2024impact}. This failure of long-term patency is driven by complex biological responses, including neointimal hyperplasia and lumen remodelling, which vary depending on the treatment approach and patient-specific factors \cite{donadoni2020multiscale, colombo2021baseline, colombo2022superficial, carbonaro2023design, corti2024impact, ninno2023systematic}. Evaluating which patients are at higher risk of restenosis, and where within the vessel the pathological changes may occur, remains a major clinical challenge.

Computational Fluid Dynamics (CFD) has been widely applied to investigate  haemodynamic characteristics in cardiovascular diseases \cite{donadoni2020multiscale, gokgol2019prediction, colombo2020computing, colombo2021baseline, colombo2022superficial, Ninno2024modelling_lower_limb, ninno2023systematic, carbonaro2023design, corti2024impact, bonfanti2017computational, Bonfanti2020, Stokes2021, Stokes2023The_Influence, stokes2023aneurysmal, hoogendoorn2020multidirectional, candreva2022current, psiuk2024methodology, de2024predicting}.
In particular, CFD studies of restenosis following PAD treatment have shown that haemodynamic indices related to Wall Shear Stress (WSS) such as Time-averaged WSS (TAWSS) and Oscillatory Shear Index (OSI) are associated with restenosis progression: regions of low TAWSS or high OSI are consistently linked to inflammation, endothelial dysfunction and tissue regrowth. \cite{donadoni2020multiscale, gokgol2019prediction, colombo2020computing, colombo2021baseline, colombo2022superficial, carbonaro2023design, corti2024impact}. 
The prominent examples include the study by G\"{o}kG\"{o}l et al. \cite{gokgol2019prediction} where CFD-derived areas with low TAWSS ($<$0.5 Pa), high TAWSS ($>$7 Pa), and high OSI ($>$0.3) were used to predict restenosis after 6 months. Donadoni et al. \cite{donadoni2020multiscale} proposed a novel haemodynamic metric called Highly Oscillatory Low Magnitude Shear (HOLMES), which integrates low TAWSS with high OSI and was shown to identify regions prone to neointimal hyperplasia. Colombo et al.\cite{colombo2021baseline} used TAWSS, OSI and Relative Residence Time (RRT) computed from CFD to identify critical regions using percentile-based thresholds, demonstrating that areas of low TAWSS are correlated with restenosis risk. Beyond restenosis in PAD, TAWSS and OSI indices have also been extensively applied in the study of coronary artery disease in a similar manner \cite{hoogendoorn2020multidirectional, candreva2022current, psiuk2024methodology, de2024predicting}. For instance, Hoogendoorn et al. \cite{hoogendoorn2020multidirectional} used critical regions based on artery-specific percentile-based thresholds of TAWSS, OSI and other haemodynamic indices to show that atherosclerotic plaque initiation and growth are associated with low and multidirectional WSS regions. Using a similar percentile-based threshold approach, De Nisco et al. \cite{de2024predicting} showed that coronary plaque progression was associated with areas that exhibit high Topological Shear Variation Index (TSVI) and low TAWSS.

Despite the great contribution of CFD to the field, its application to routine clinical practice is limited due to its high computational cost, long processing times, reliance on proprietary
software, and the need for expert judgment to balance the trade-off between accuracy and complexity \cite{BONFANTI2018}. In response to these limitaions, data-driven approaches, which include various types of machine learning (ML) and reduced-order modelling (ROM), have increasingly been adopted alongside traditional CFD. These methods have shown effectiveness across a range of haemodynamic studies, ranging from the prediction of blood haemodynamic quantities \cite{Liang2019A, du2022deep, yao2024image2flow, alamir2024rapid, wong2024strategies, chan2025role, pajaziti2023shape, drakoulas2023fastsvd, macraild2024reduced, chatpattanasiri2025ML, barzegar2025predictive, siena2023data}, to the enhancement of data resolution and noise reduction \cite{chatpattanasiri2023towards, Ferdian20204DFlowNet, fathi2020super, kontogiannis2022joint, csala2024comparison}. 

The key strength of ML lies in its ability to extract and exploit complex patterns that exist in large datasets \cite{Shlezinger2020Model, alpaydin2020introduction_to_ML}. However, higher data dimensionality often leads to greater computational demands and increased model sensitivity to noise \cite{alpaydin2020introduction_to_ML}. Dimensionality reduction helps address both issues. \cite{alpaydin2020introduction_to_ML, Arzani2021, Brunton_Kutz_2022}. Proper Orthogonal Decomposition (POD), also known as Principal Component Analysis (PCA), is one of the most commonly used dimensionality reduction techniques \cite{chatpattanasiri2023towards, Du2018Dimensionality, chatpattanasiri2025ML}. POD works by applying Singular Value Decomposition (SVD) to extract a set of orthogonal modes. The original high-dimensional Full Order Model (FOM) is then projected onto this reduced set of modes, yielding a compact Reduced Order Model (ROM). The ROM flow field is represented by coefficients that quantify each mode’s contribution, providing an efficient low-dimensional description while preserving essential physics. POD has been applied across cardiovascular settings to reduce high-dimensional flow data. For example, Di Labbio
and Kadem \cite{di2019reduced} investigated coherent flow structures in a left ventricle with aortic regurgitation, comparing POD with Dynamic Mode Decomposition (DMD), and Chatpattanasiri et al.  \cite{chatpattanasiri2023towards} applied an extension of POD called Robust POD to construct computationally efficient ROMs of the velocity field in a dissected aortic dissection.

Another powerful use of ROMs is in combination with ML models for predictive tasks. In such integrated POD–ML frameworks, ML models are trained to map input parameters (such as boundary conditions and vascular geometries) to POD coefficients, which are then used to reconstruct haemodynamic fields of interest \cite{pajaziti2023shape, drakoulas2023fastsvd, macraild2024reduced, barzegar2025predictive, chatpattanasiri2025ML}. For instance, Pajaziti et al. \cite{pajaziti2023shape} trained Feed-forward Neural Networks (FFNNs) to predict PCA coefficients of velocity and pressure fields in an aorta based on the PCA coefficients of the shapes of each aorta, and Chatpattanasiri et al. \cite{chatpattanasiri2025ML} developed a POD–ML framework that uses LSTMs and FFNNs to predict time-dependent WSS directly from the inlet flowrate waveform under a limited training data regime, demonstrating the performance through two case studies in patient-specific arteries: a femoral artery and a dissected aorta. More recently, Gerdroodbary and Salavatidezfouli \cite{barzegar2025predictive} utilised POD and LSTM networks to predict velocity and pressure in patient-specific carotid arteries with and without stenosis, evaluating reconstruction errors across different POD mode numbers. 

However, existing ML-ROM haemodynamic studies have typically focused on one source of variability at a time. Several works explored steady-state or a fixed inlet boundary condition flow fields across different vascular geometries \cite{Liang2019A, du2022deep, pajaziti2023shape, yao2024image2flow, alamir2024rapid, wong2024strategies, chan2025role}. Others investigated unsteady flows in a fixed geometry under varying inlet flowrate waveforms \cite{drakoulas2023fastsvd, macraild2024reduced, chatpattanasiri2025ML, barzegar2025predictive}. To the best of the authors’ knowledge, no study has yet combined both aspects in the same model, which is essential to capture the full spectrum of haemodynamic variability. This gap is due to the substantial complexity needed for an ML model to account for both sources of variability simultaneously. Each new geometry already alters local haemodynamics in a nonlinear manner, while waveform variation adds further temporal dynamics that interact with these geometric effects. Capturing this joint dependence may require a large, high-fidelity dataset and an overly complex model capable of generalising across diverse spatio-temporal patterns. Siena et al. \cite{siena2023data} attempted a step in this direction by examining geometric and inflow variability within one ML-ROM prediction framework. However, they explored these sources of variability separately as stenosis severity was varied with fixed inflow conditions, while inflow scaling was studied with fixed stenosis severity.

In our work, we advance the state of the art by introducing a new ML-ROM framework that accounts for variability in \textit{both} vessel geometry and inlet flowrate waveform simultaneously. We generated a set of synthetic of geometries by deforming a baseline model using six geometric parameters related to the tapering, stenosis, and curvature. In parallel, synthetic inlet flowrate waveforms were created by applying PCA to a set of real patient waveforms, and then randomising the PCA coefficients to capture physiologically realistic variation. To keep the computational demand tractable, rather than directly attempting to predict full TAWSS and OSI distributions, we post-processed the CFD results to compute haemodynamic indices (TAWSS and OSI), and applied thresholds reported in prior studies to identify the critical regions (regions on the vessel's surface that are associated with risk of restenosis). POD was then used to construct ROMs of the critical regions, and ML models were trained to predict the corresponding ROM coefficients. Finally, the predicted critical regions are obtained by reconstructing the ROMs with these predicted coefficients. This framework efficiently incorporates both anatomical and waveform variability while remaining computationally tractable.

\section{Methods}

\begin{figure}[ht]
\centering
\includegraphics[width=\textwidth]{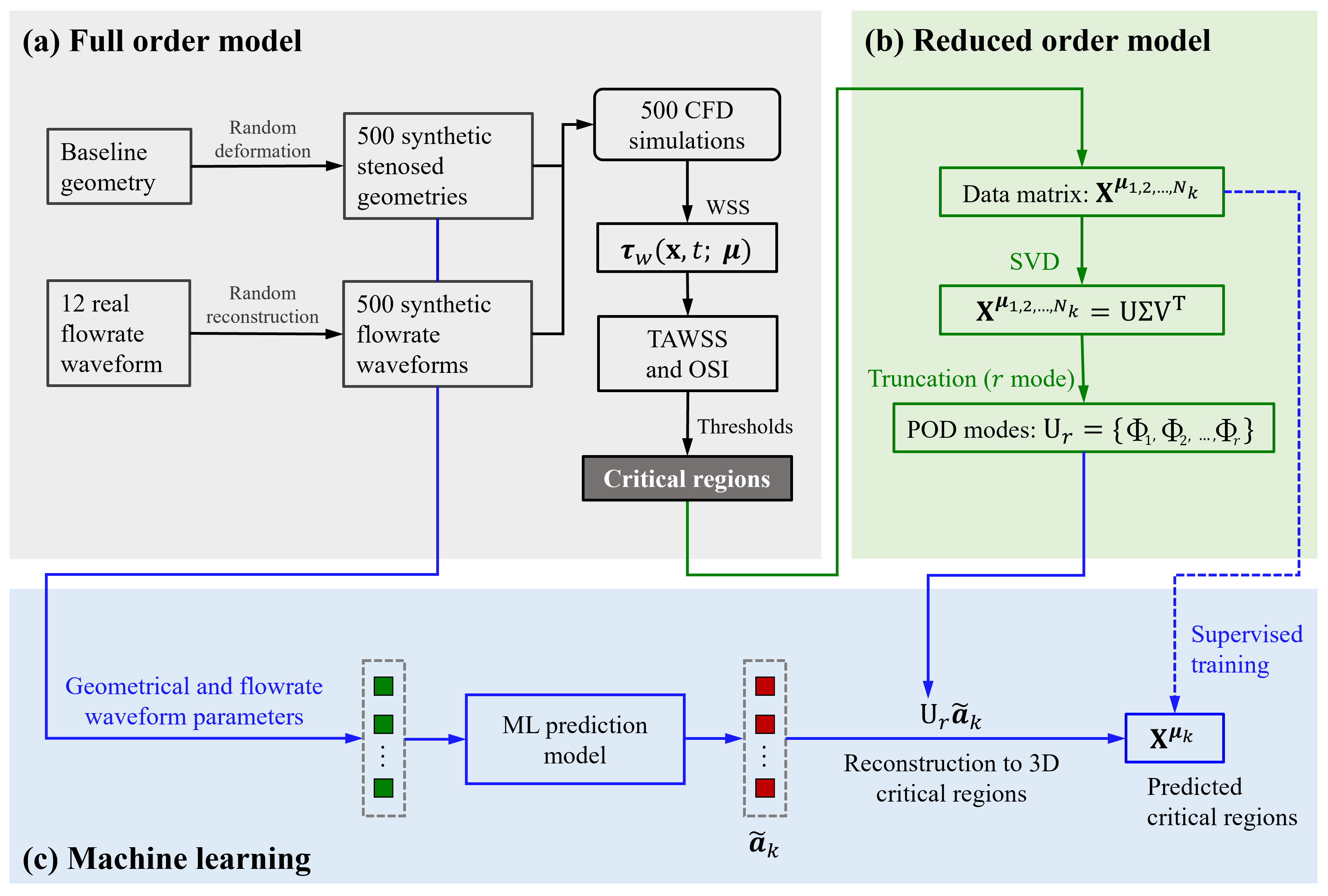}
\caption{
Diagrammatic overview of the study methodology. (a) Synthetic stenosed geometries and inflow waveforms were generated and used in CFD simulations to obtain haemodynamic indices, and thresholds were applied to identify critical regions which served as FOMs. (b) POD was applied to FOMs to identify POD modes. (c) ML models were trained on geometric and inlet flowrate waveform parameters to predict POD coefficients, which were then used to reconstruct the predicted critical regions.}
\label{fig:overall_process}
\end{figure}

Figure \ref{fig:overall_process} provides an overview of the workflow of this study. The process begins with the FOM step, Figure \ref{fig:overall_process}a) where a baseline femoral artery geometry was parametrically deformed to generate 500 synthetic stenosed geometries, while 12 real flowrate waveforms were used to reconstruct 500 physiologically realistic inflow waveforms. CFD was used to simulate the flow in each geometry–waveform pair. WSS distributions were then calculated from which TAWSS and OSI were derived. Next, critical regions associated with restenosis risk were identified by applying threshold-based criteria: two artery-specific percentile-based thresholds, TAWSS33 (TAWSS $<33^{\text{rd}}$ percentile) and OSI66 (OSI $>66^{\text{th}}$ percentile), and two absolute thresholds based on reported values in the literature \cite{ninno2023systematic}, TAWSS0.5 (TAWSS $<0.5$ Pa) and OSI0.2 (OSI $>0.2$). This led to four distinct FOMs, one for each threshold, which were handled separately in the subsequent ROM and ML stages. Further details on dataset generation can be found in Section \ref{ssec:Dataset generation}.
In the ROM step (Figure \ref{fig:overall_process}b), POD was applied to each of the FOMs to extract a set of orthogonal modes that efficiently represent the spatial distribution of the critical regions while retaining their essential features as explained in Section \ref{ssec:POD}. Lastly, Figure \ref{fig:overall_process}c shows the supervised ML models, which were trained on geometric and inlet flowrate waveform parameters. The models predicted the reduced-order representations (can be viewed as POD coefficients), which were subsequently mapped back onto the POD modes to reconstruct the critical regions. More details on the ML models, including the ML architecture explored, training workflow, and hyperparameter tuning strategy, are presented in Section \ref{ssec:ML models}.

\subsection{Dataset generation} \label{ssec:Dataset generation}

\begin{figure}
\centering
\includegraphics[width=\textwidth]{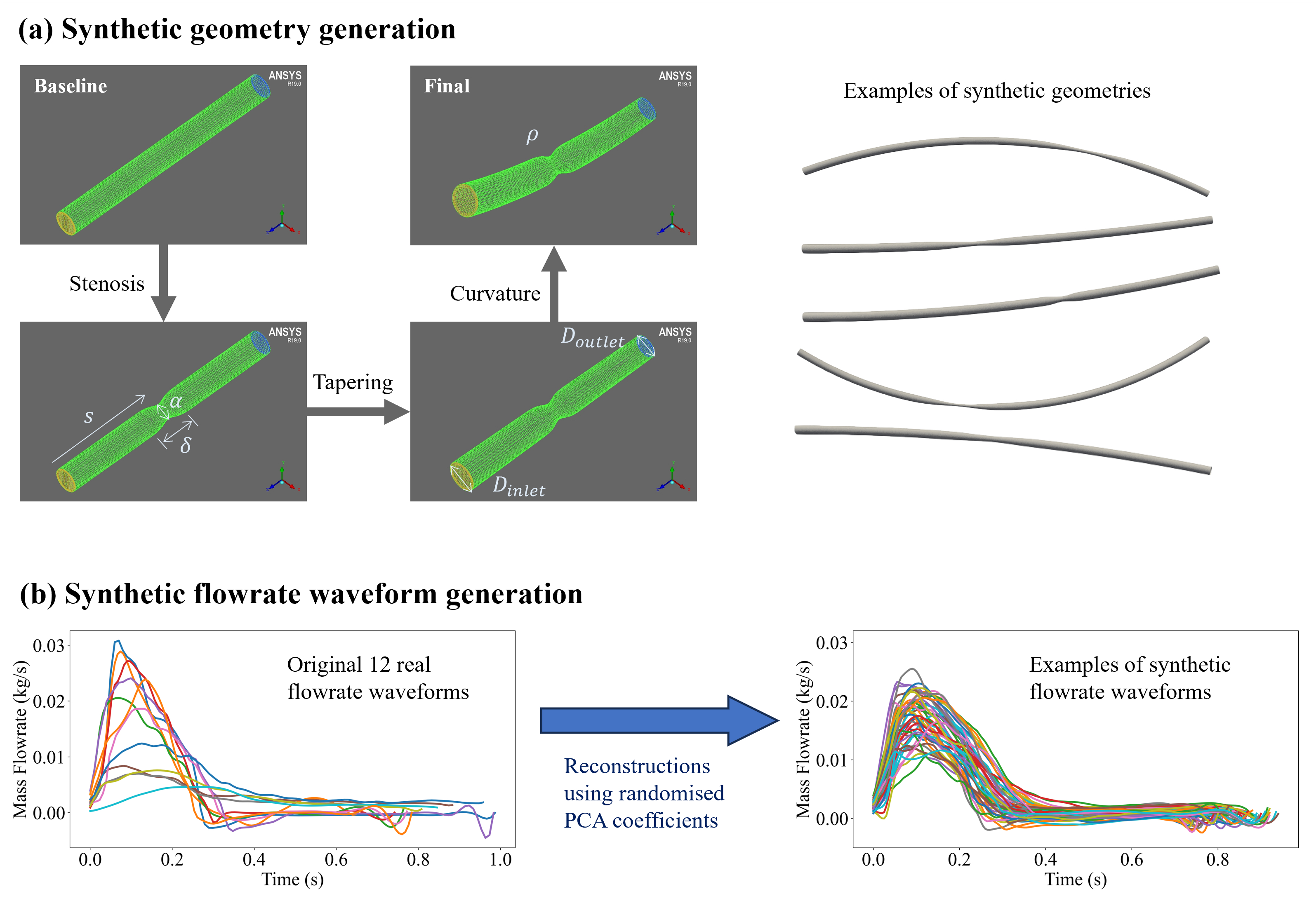}
\caption{
Dataset generation. (a) Synthetic geometries are created by deforming a baseline model in three steps, using six geometric parameters. Examples of generated geometries are shown. (b) Synthetic flowrate waveforms obtained by applying PCA to 12 real flowrate waveforms and randomising coefficients. Examples of original and synthetic waveforms are shown.}
\label{fig:dataset_generation}
\end{figure}

\begin{table}[ht]
\centering
\caption{Definition and sampling ranges of geometric parameters used for synthetic geometry generation.}
\label{tab:geo_parameters}
\begin{tabular}{llll}
\hline
\textbf{Parameter} & \textbf{Symbol} & \textbf{Definition} & \textbf{Range} \\
\hline
Inlet diameter      & $D_{\text{inlet}}$ & Vessel diameter at inlet cross-section          & 4.8 - 7.2 mm \\
Outlet diameter     & $D_{\text{outlet}}$ & Vessel diameter at outlet cross-section        & 4.3 - 6.5 mm \\
Stenosis severity   & $\alpha$     & Maximum ratio of reduction in diameter         & 0.2 - 0.7 \\
Stenosis length     & $\delta$     & The length of stenosis                         & 14.4 - 66.4 mm \\
Stenosis position   & $s$          & Axial location of stenosis centre              & 120 - 210 mm \\
Curvature           & $\rho$       & Vessel curvature (inverse radius)              & -0.004 - 0.004 mm$^{-1}$ \\
\hline
\end{tabular}
\end{table}

This section provides a brief overview of the dataset generation process. Full methodological details are reported in Supplementary Material 1. Figure \ref{fig:dataset_generation} illustrate the overall process.

The dataset generation began with a baseline model representing a straight cylindrical vessel, constructed with diameter and length values characteristic of the human femoral artery. This model was meshed and validated for a mesh independence test before serving as the template for subsequent deformations. Geometric variability was introduced through six parameters: inlet ($D_{\text{inlet}}$) and outlet diameters ($D_{\text{outlet}}$), stenosis severity ($\alpha$), stenosis length ($\delta$), stenosis position ($s$), and overall curvature ($\rho$), with their definitions and ranges summarised in Table \ref{tab:geo_parameters}. Parameter values were sampled using Latin Hypercube Sampling (LHS) to ensure efficient and uniform coverage of the multidimensional space. Each baseline mesh was then systematically deformed in three steps. First, a localised Gaussian-shaped stenosis was imposed to control narrowing severity and length (similar to studies by Buoso et al. \cite{buoso2019reduced} and Wong et al. \cite{wong2024strategies}). Second, a tapering transformation adjusted the diameter variation between the inlet and outlet. Finally, a curvature transformation introduced uniform vessel bending. This procedure generated 500 synthetic stenosed geometries spanning a wide and physiologically realistic spectrum of lumen narrowing and tortuosity, as illustrated in Figure \ref{fig:dataset_generation}a.

To generate 500 synthetic inlet flowrate waveforms, a set of twelve patient-derived femoral artery inflow waveforms was first collected to establish the base population, shown in Figure \ref{fig:dataset_generation}b (left). PCA was applied to these real waveform signals, representing them using a set of twelve PCA coefficients. To synthesise new waveforms, each coefficient was sampled using Latin Hypercube Sampling (LHS) within a bounded interval defined by its mean $\pm$ one standard deviation across the patient dataset, ensuring realistic yet diverse variations. In addition, the waveform period was treated as a separate parameter and sampled using LHS in a similar manner. Each synthetic waveform was reconstructed from the sampled coefficients and rescaled to the sampled period before being uniformly interpolated onto a fixed temporal grid with step size $\Delta t = 0.005$ s. The resulting synthetic population (Figure \ref{fig:dataset_generation}b, right) provided 500 distinct inflows that retained clinical realism while spanning a wide range of physiologically plausible dynamics. 

To confirm physiological relevance, Reynolds and Womersley numbers were computed using the inlet diameter $D_{\text{inlet}}$, and assuming a value of blood viscosity of $\mu = 0.004~\mathrm{Pa\cdot s}$ for all cases \cite{nader2019blood}. The Reynolds number based on time-averaged flow rate spans $95.5$ to $270.0$, while the peak Reynolds number (using the cycle-maximum flow rate) spans between $287.7$ and $1026.0$. The Womersley number, also using $D_{\text{inlet}}$, varies from $6.52$ to $10.72$. These ranges are consistent with values reported in the literature for femoral arteries \cite{huo2018acute, williamson2024literature}.

Blood flow in each geometry–waveform pair was simulated using CFD implemented in Ansys Fluent 19.0 (Ansys Inc., PA, USA). Blood was modelled as a non-Newtonian fluid with Carreau viscosity, and a time-dependent parabolic velocity profile matching the synthetic waveform was prescribed at the inlet. Simulations were run for two cardiac cycles, with the first cycle discarded to remove transient effects. The 500 simulation results were split into 300:100:100 for training, validation, and testing, respectively, for the subsequent ROM (Section \ref{ssec:POD}) and ML (Section \ref{ssec:ML models}) phases. To validate the reliability of the CFD setup, additional simulations were performed by replicating the experimental and numerical conditions of cases studied by Hong et al. \cite{hong2017characteristics}, who investigated pulsatile flow in curved and stenosed channels. Comparison of velocity fields confirmed good agreement and provided confidence in our simulations.

WSS distributions were extracted and used to compute TAWSS and OSI according to Equations\eqref{eq:TAWSS_and_OSI}:

\begin{subequations} 
\begin{align} 
\text{TAWSS} &= \frac{1}{T} \int_{0}^{T} |\boldsymbol{\tau}_w| \, dt \label{eq:TAWSS} 
\\ \text{OSI} &= 0.5 \left(1 - \frac{|\int_{0}^{T} \boldsymbol{\tau}_w \, dt|}{\int_{0}^{T} |\boldsymbol{\tau}_w| \, dt}\right) \label{eq:OSI} 
\end{align} \label{eq:TAWSS_and_OSI} 
\end{subequations}

where $\boldsymbol{\tau}_w$ is the instantaneous WSS vector and $T$ is the cardiac cycle period.

Critical regions were then defined by applying threshold-based criteria to the TAWSS and OSI distributions. Although numerous studies agree that low TAWSS and high OSI are associated with restenosis risk, the thresholds used in the literature vary considerably \cite{hoogendoorn2020multidirectional, Ninno2024modelling_lower_limb, ninno2025silico, de2024predicting, ninno2023systematic}. For this reason, our aim was not to determine the optimal threshold, but rather to demonstrate that our ML framework can robustly predict critical regions under different plausible definitions. To this end, we considered four types of critical-region definitions based on two types of thresholding approaches. The first approach applied artery-specific (or relative) thresholds: TAWSS33 for TAWSS values below the 33rd percentile and OSI66 for OSI values above the 66th percentile. This percentile-based approach has been used in multiple studies involving femoral, coronary and other arteries \cite{hoogendoorn2020multidirectional, Ninno2024modelling_lower_limb, ninno2025silico, de2024predicting}. The second approach applied absolute thresholds: TAWSS0.5 (TAWSS $<0.5$ Pa) and OSI0.2 (OSI $>0.2$), reported as commonly adopted thresholds in Ninno et al. \cite{ninno2023systematic}. Each threshold produced a binary representation of critical regions, which served as a separate FOM for the subsequent phases.

\subsection{POD} \label{ssec:POD}

POD was used to obtain low-dimensional bases for the critical-region distributions. For each thresholding criterion, a separate ROM was constructed. Only a brief description of POD is provided here, and more details can be found in \cite{Brunton_Kutz_2022, Berkooz1993ThePOD}.

Let $N$ be the number of vessel-wall nodes (identical number across cases by construction of the deformed baseline mesh). For case $\mu_k$, let $\mathbf{x}^{(q)}_k\in\{0,1\}^{N}$ denote a binary vector indicating whether each node belongs to the critical region $q \in \{\text{TAWSS33}, \text{OSI66}, \text{TAWSS0.5}, \text{OSI0.2}\}$. Stacking the mean-centred $N_{\text{train}}$ training cases gives the data matrix

\begin{equation}
X^{(q)} =
\big[\, \mathbf{x}_1 - \bar{\mathbf{x}}^{(q)},\ 
       \mathbf{x}_2 - \bar{\mathbf{x}}^{(q)},\ \ldots,\ 
       \mathbf{x}_{N_{\text{train}}} - \bar{\mathbf{x}}^{(q)} \,\big],
\end{equation}

with $\bar{\mathbf{x}}^{(q)}$ denoting the mean values across the training dataset. SVD is then applied to $X^{(q)}$:

\begin{equation}
X^{(q)} = U^{(q)} \Sigma^{(q)} {V^{(q)}}^\top . 
\end{equation}

The columns of $U^{(q)}$ are the POD spatial modes $\{\Phi_i^{(q)}\}$ ordered by decreasing singular values $\{\sigma_i^{(q)}\}$. Truncating to $r$ modes yields the basis $U^{(q)}_r = [\Phi^{(q)}_1,\Phi^{(q)}_2,\ldots,\Phi^{(q)}_r]$.

For any case $\mu_k$, the POD coefficients are

\begin{equation}
\mathbf{a}^{(q)}_k = {U^{(q)}_r}^\top\!\left(\mathbf{x}^{(q)}_k - \bar{\mathbf{x}}^{(q)}\right),
\end{equation}

and the $r$-mode reconstruction is

\begin{equation}
\hat{\mathbf{x}}^{(q)}_k = \bar{\mathbf{x}}^{(q)} + U^{(q)}_r\, \mathbf{a}^{(q)}_k.
\end{equation}

$\hat{\mathbf{x}}^{(q)}_k$ can then be re-binarised by another threshold value. The reconstruction performance was assessed by framing each vessel surface as a binary classification problem, where predicted and true critical-region distributions were compared node-wise to compute true positives (TP), true negatives (TN), false positives (FP), and false negatives (FN). To account for potential class imbalance, the reconstruction performance was evaluated using balanced accuracy (BA), defined as the mean of \textit{sensitivity} and \textit{specificity}:

\begin{equation}
\text{Sensitivity} = \frac{\text{TP}}{\text{TP}+\text{FN}}, \qquad
\text{Specificity} = \frac{\text{TN}}{\text{TN}+\text{FP}}
\end{equation}

\begin{equation}
\text{BA} = \tfrac{1}{2}\big(\text{Sensitivity} + \text{Specificity}\big)
= \tfrac{1}{2}\left(\frac{\text{TP}}{\text{TP}+\text{FN}} + \frac{\text{TN}}{\text{TN}+\text{FP}}\right).
\label{eq:BA}
\end{equation}

\subsection{ML models} \label{ssec:ML models}

\begin{figure}
\centering
\includegraphics[width=0.8\textwidth]{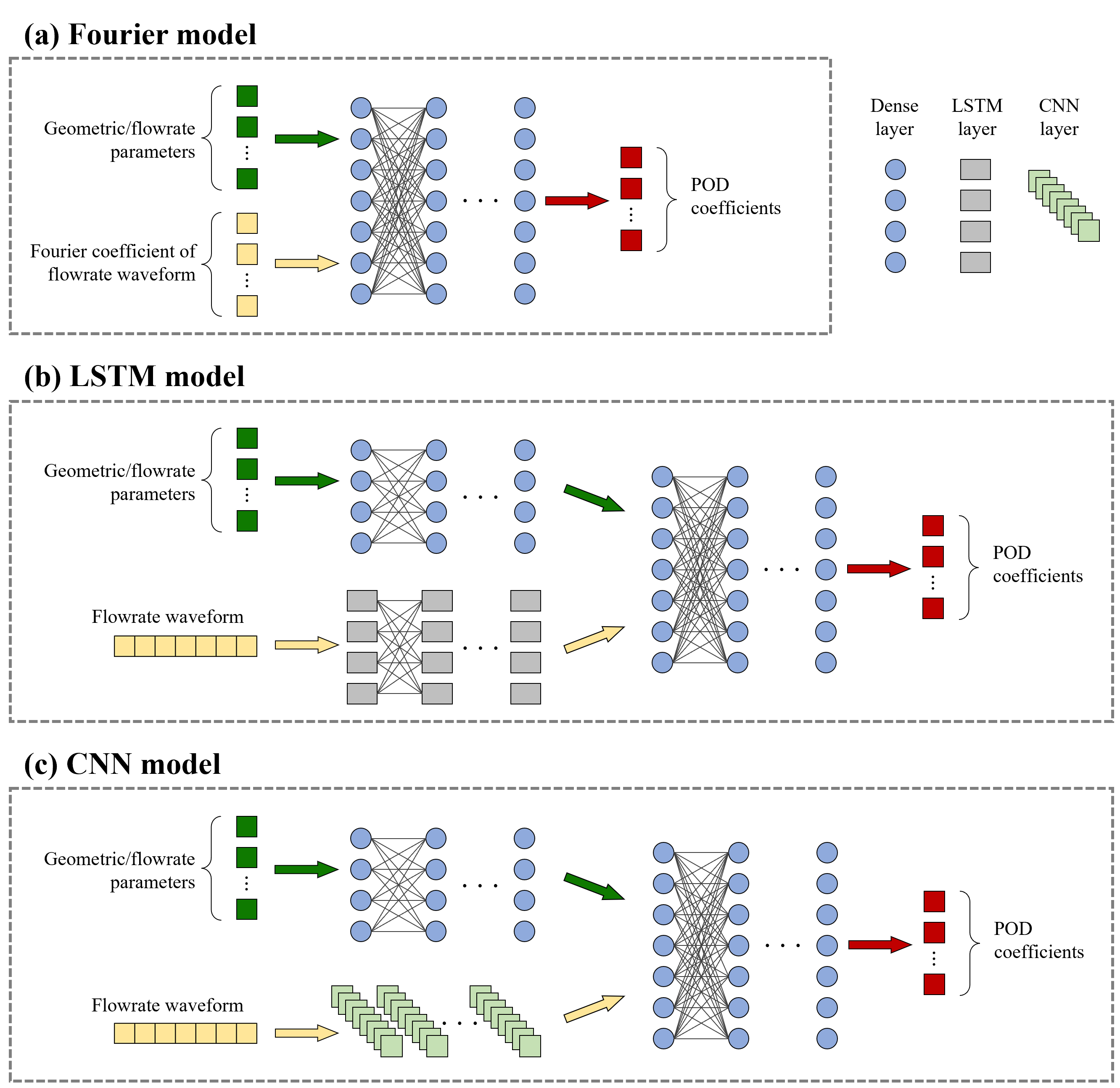}
\caption{
ML models. The models explored in our study are named after their treatment of the inlet flowrate waveform input: (a) Fourier model, (b) LSTM model, and (c) CNN model.}
\label{fig:ML_models}
\end{figure}

Three ML models were explored to map inputs to the POD coefficients of the critical-region distributions, each predicting an $r$-dimensional coefficient vector that was subsequently used to reconstruct the regions. The models were named after their treatment of the flowrate waveform: the Fourier model represented it by Fourier coefficients, while the LSTM and CNN models learned directly from the waveform time series sampled at 100 time intervals (Figure \ref{fig:ML_models}). All models took as input two groups of features: the six geometric parameters appended with two scalar flowrate parameters (average flowrate and period), and the waveform representation specific to each model.

\begin{enumerate}
\item \textbf{Fourier model.}
The waveform was represented by its Fourier coefficients: sine ($a_i$) and cosine ($b_i$) which were concatenated with the geometric and scalar waveform features to form a single vector. This vector was then passed through multiple dense layers with Rectified Linear Unit (ReLU) activations to map to the POD coefficients.

\item \textbf{LSTM model.}
The waveform sequence was fed into stacked LSTM layers, while the geometric/flowrate parameters were passed through dense layers. The two embeddings were concatenated and then mapped through dense layers to predict the POD coefficients. 

\item \textbf{CNN model.}
The waveform sequence was processed using 1D convolution layers with max-pooling, then aggregated via global average pooling. The geometric/flowrate features were handled using dense layers. The results were concatenated and passed through another set of dense layers to predict the POD coefficients.
\end{enumerate}

The Adam optimiser was employed for training. The training process was set to automatically terminate if the validation loss did not decrease for 20 consecutive epochs, with the model parameters restored to their values 20 epochs earlier. The loss function was defined as a class-balanced binary cross-entropy, where the weights were computed as the ratio of class imbalance between positive and negative samples in each batch:

\begin{equation}
\mathcal{L} = -\frac{1}{NN_b} \sum_{i=1}^{N} \sum_{j=1}^{N_b} \Bigg[
w_{\mathrm{pos}}\, x^{(q)}_{i,j} \log\!\left(\tilde{x}^{(q)}_{i,j} + \varepsilon\right)
+ w_{\mathrm{neg}} \left(1 - x^{(q)}_{i,j}\right) 
\log\!\left(1 - \tilde{x}^{(q)}_{i,j} + \varepsilon\right) \Bigg],
\end{equation}

where $x^{(q)}_{i,j}$ is the ground truth critical-region label (0 or 1) for node $i$, case $j$, $\tilde{x}^{(q)}_{i,j}$ is the predicted value after reconstruction by POD, $w_{\mathrm{pos}}$ and $w_{\mathrm{neg}}$ are class-balancing weights, $N_b$ is the batch size, and $\varepsilon=10^{-6}$ is a small constant for numerical stability\footnote{Specifically for cases where $\tilde{x}^{(q)}_{i,j}$ is close to 0 or 1.}.

To obtain the loss, the predicted coefficients $\tilde{\mathbf{x}}^{(q)}_k$ were first mapped back to a critical-region field $\tilde{\mathbf{x}}^{(q)}_k=\bar{\mathbf{x}}^{(q)}+U^{(q)}_r \tilde{\mathbf{a}}^{(q)}_k$, followed by a steep sigmoid to provide a soft binarisation relative to the chosen threshold. Hyperparameters, including parameters related to the model architecture, regularisation coefficients, learning rate, the number of POD coefficients, and threshold values, were selected automatically using Optuna\cite{optuna_2019}. In each Optuna trial, ML training was repeated 10 times with different initialisations; the trial score with the highest mean BA on the validation dataset was taken. After 200 trials, the best model was re-evaluated on the test dataset.

\section{Results} \label{sec:Results}

\subsection{POD reconstruction performance} \label{ssec:POD reconstruction performance}

\begin{figure}
\centering
\includegraphics[width=0.70\textwidth]{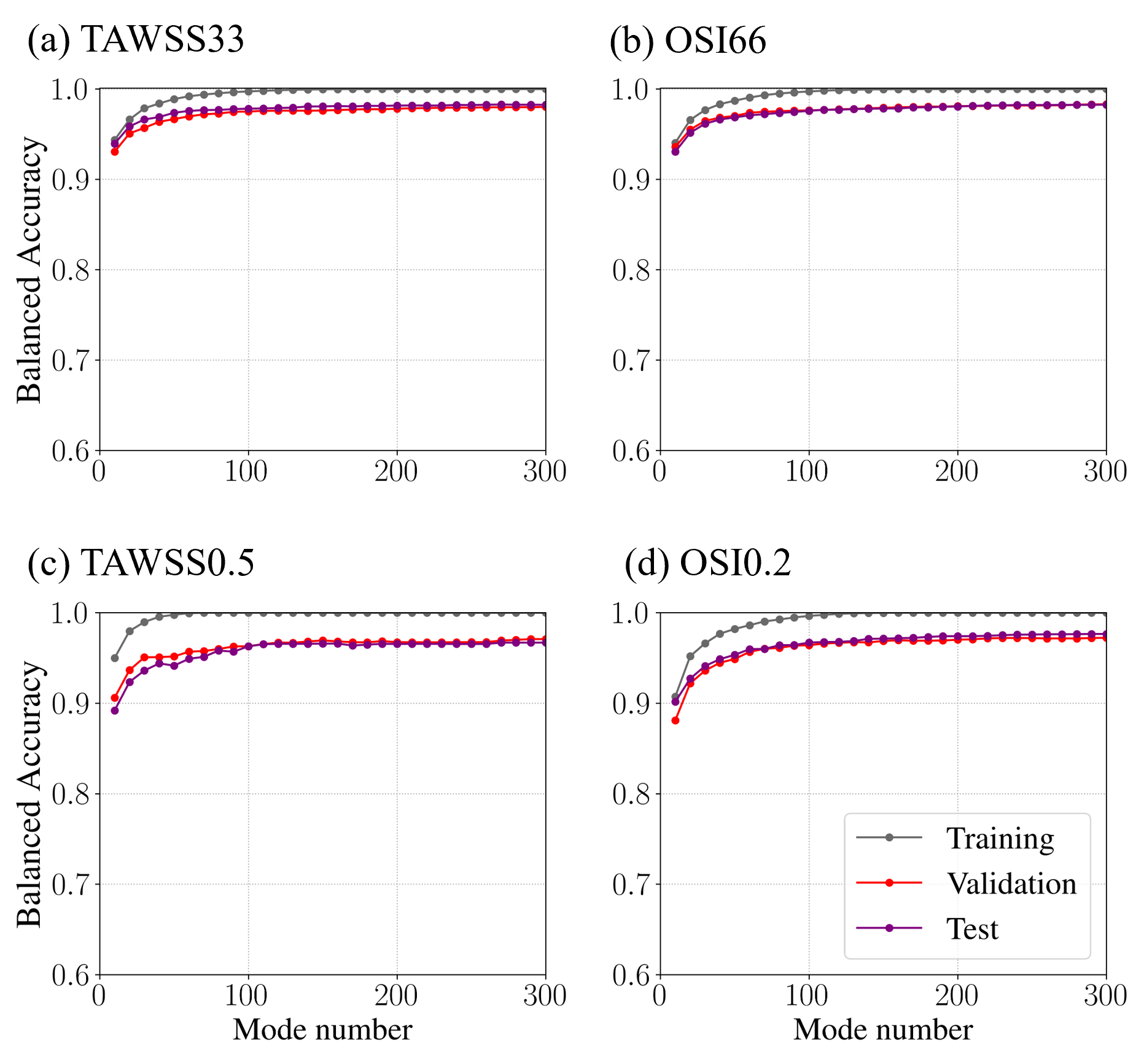}
\caption{
POD reconstruction performance. BA of reconstructed critical regions as a function of mode number $i$ for (a) TAWSS33, (b) OSI66, (c) TAWSS0.5, and (d) OSI0.2. Results are shown for training (grey), validation (red), and test (purple) datasets.}
\label{fig:POD_recon}
\end{figure}

The reconstruction performance of the POD was evaluated for each of the four critical-region definitions. BA was used to quantify reconstruction quality, comparing the original and reconstructed critical-region data across all vessel-wall nodes. Figure \ref{fig:POD_recon} presents the median BA values as a function of the number of retained modes for training, validation, and test datasets.

Reconstructing critical regions based on relative thresholds (TAWSS33 and OSI66) showed rapid convergence of BA, with the value exceeding 0.95 using only 20 modes and plateauing for higher mode numbers. Convergence was slightly slower for regions based on absolute thresholds (TAWSS0.5 and OSI0.2), requiring about 50 modes to achieve similar levels of performance and exhibiting larger gaps between training and validation/test datasets compared to TAWSS33 and OSI66. This could be due to the fact that absolute thresholds produce strongly imbalanced classes, with the majority of nodes on the vessel surface identified as non-critical: while TAWSS33 and OSI66 each label 33\% of nodes as critical for all cases by definition, TAWSS0.5 identifies only 7.5\% of nodes, and OSI0.2 identifies only 17.3\% as critical on average across cases in the training dataset. These imbalances also vary across vessels, making the binary fields less homogeneous and harder for low-rank bases to capture. Nevertheless, the median BA curves for validation and test datasets remained consistently high and only slightly below the training curves, indicating that the POD bases generalised well to unseen cases.

\subsection{ML prediction performance}

\begin{figure}
\centering
\includegraphics[width=0.70\textwidth]{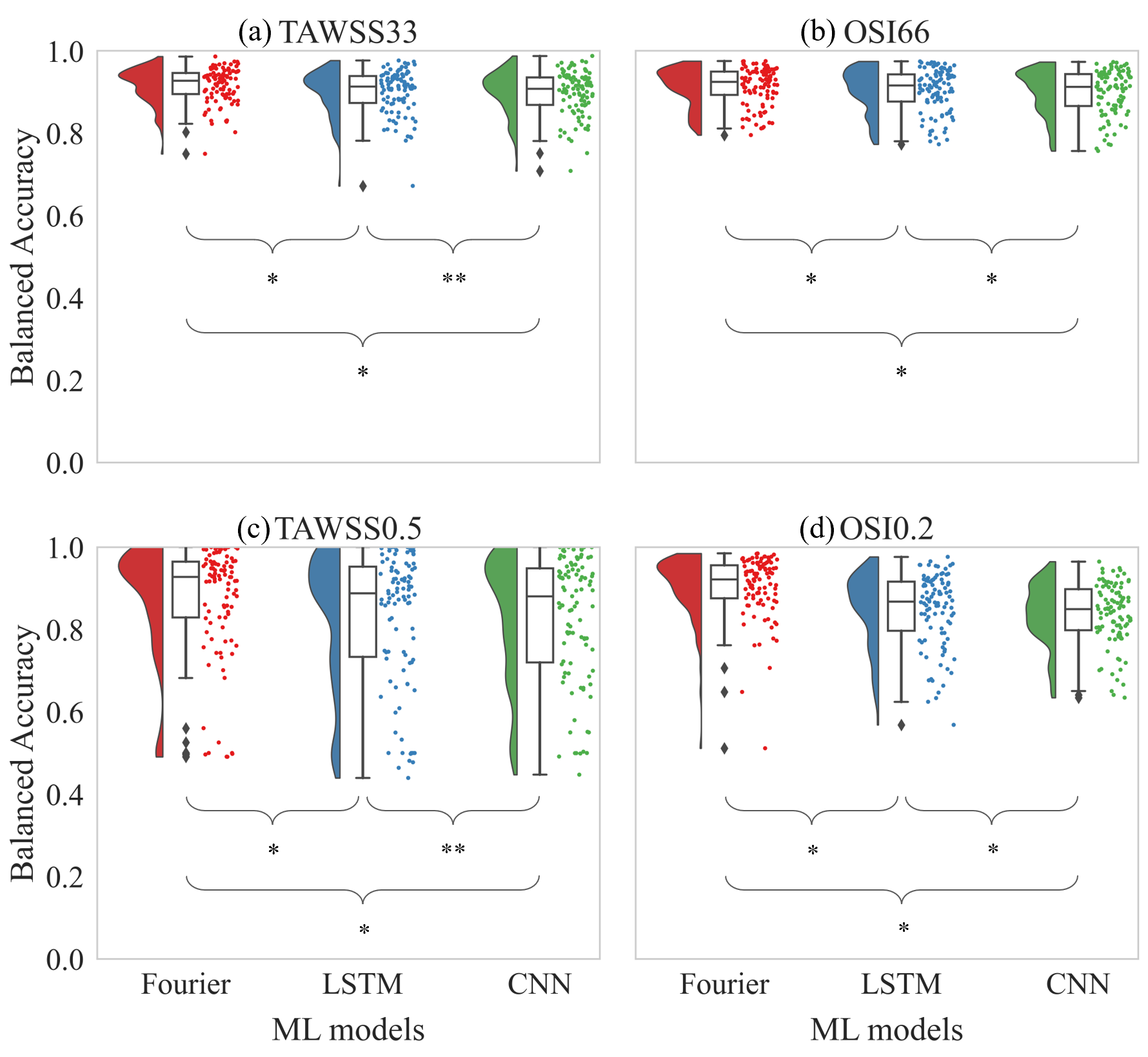}
\caption{
Comparison of test dataset BA across the three ML models for four critical-region definitions: (a) TAWSS33, (b) OSI66, (c) TAWSS0.5, and (d) OSI0.2. Each raincloud plot shows case-wise distribution, combining a half-violin (distribution), a boxplot (median and interquartile range), and a dot plot (each test case). Results of pairwise Wilcoxon signed-rank tests are shown with * for $p < 0.05$, and ** for $p \geq 0.05$ .}
\label{fig:ML_performance_compare}
\end{figure}

Figure \ref{fig:ML_performance_compare} presents BA distributions across the test dataset for the three ML architectures and four critical-region definitions. These are the models with optimal hyperparameters found by Optuna (see Supplementary Material 2 for the hyperparameters and training curves of all models). The Fourier models consistently achieved the highest performance, with median (and interquartile range, IQR) BA values of 0.928 (0.0514), 0.925 (0.0564), 0.927 (0.135), and 0.922 (0.808) for TAWSS33, OSI66, TAWSS0.5, and OSI0.2, respectively. The LSTM models yielded slightly lower performance, with median BA values between 0.868 and 0.916, while the CNN models showed the lowest performance, with median BA varying from 0.850 to 0.913. Wilcoxon signed-rank tests confirmed that Fourier models significantly outperformed both LSTM and CNN models across all four critical-region definitions ($p < 0.05$ in every case). Comparisons between LSTM and CNN revealed mixed outcomes: LSTM achieved significantly higher accuracy for OSI66 and OSI0.2, while no significant differences were observed for TAWSS33 or TAWSS0.5. The superior performance of the Fourier models may arise from their compact representation of inflow waveforms, which efficiently encodes dominant features while reducing susceptibility to overfitting compared to LSTM and CNN models. 

For all models explored, critical regions defined by relative thresholds (TAWSS33 and OSI66) produced more stable predictions, with IQRs ranging between 0.0514 and 0.0772 across models (e.g., Fourier: 0.0514 for TAWSS33 and 0.0564 for OSI66; LSTM: 0.0652 and 0.0659; CNN: 0.0663 and 0.0772). In contrast, critical regions defined by absolute thresholds (TAWSS0.5 and OSI0.2) were associated with wider variability, with IQRs extending from 0.0808 up to 0.228 (e.g., Fourier: 0.135 for TAWSS0.5 and 0.0808 for OSI0.2; LSTM: 0.219 and 0.119; CNN: 0.228 and 0.100).

For brevity, subsequent analyses focused exclusively on the results of the Fourier models, since the models of this architecture achieved the highest performance across all critical-region definitions.


\begin{figure}
\centering
\includegraphics[width=0.70\textwidth]{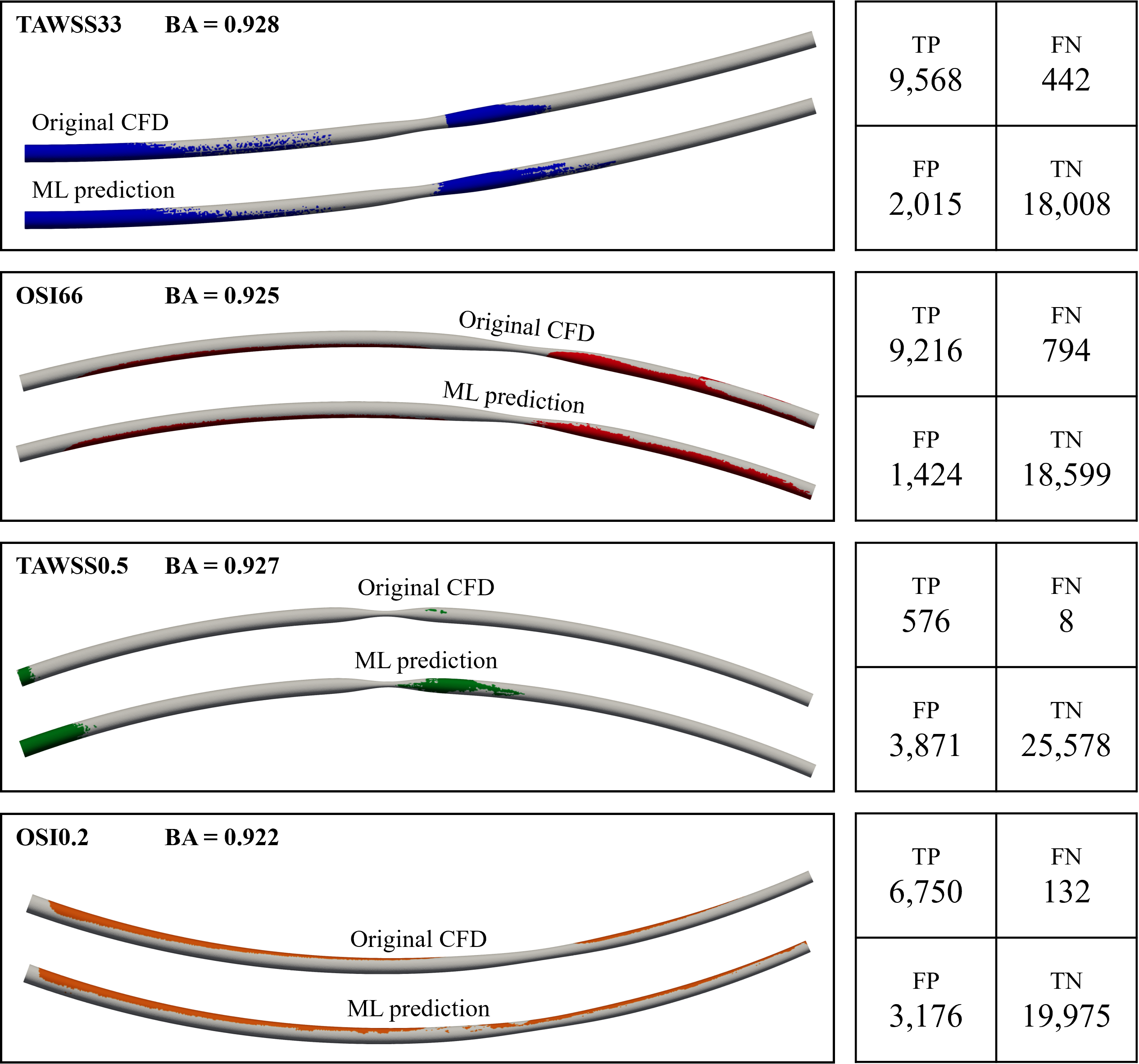}
\caption{
Three-dimensional reconstructed critical region of the median case for each definition using Fourier models, comparing ML-predicted regions with the original CFD-derived ones. The corresponding confusion matrix for each case is shown on the right.}
\label{fig:3D_Reconstructions}
\end{figure}

To qualitatively assess prediction accuracy, Figure \ref{fig:3D_Reconstructions} presents examples of reconstructed critical regions predicted by the Fourier models. The corresponding confusion matrix for each is shown on the right. These reconstruction examples are the median-performing cases in the test dataset. Our Fourier-based framework correctly predicted critical-region distributions for all four critical-region definitions, with the median BA values exceeding 0.92. Minor discrepancies were primarily observed around the critical regions, where threshold-based binarisation is prone to errors due to small variations. It should be noted that even for the case of TAWSS0.5 in which many nodes are misclassified as critical (high FP), the BA value is still high. This is due to the model identifying nearly all the critical nodes correctly (high TP), resulting in high sensitivity. At the same time, TN greatly outnumbers FP, hence the specificity remains acceptably high. This explains the high value of BA in TAWSS0.5 (see equation \ref{eq:BA}).



\subsection{Robustness analysis}

\begin{figure}
\centering
\includegraphics[width=0.9\textwidth]{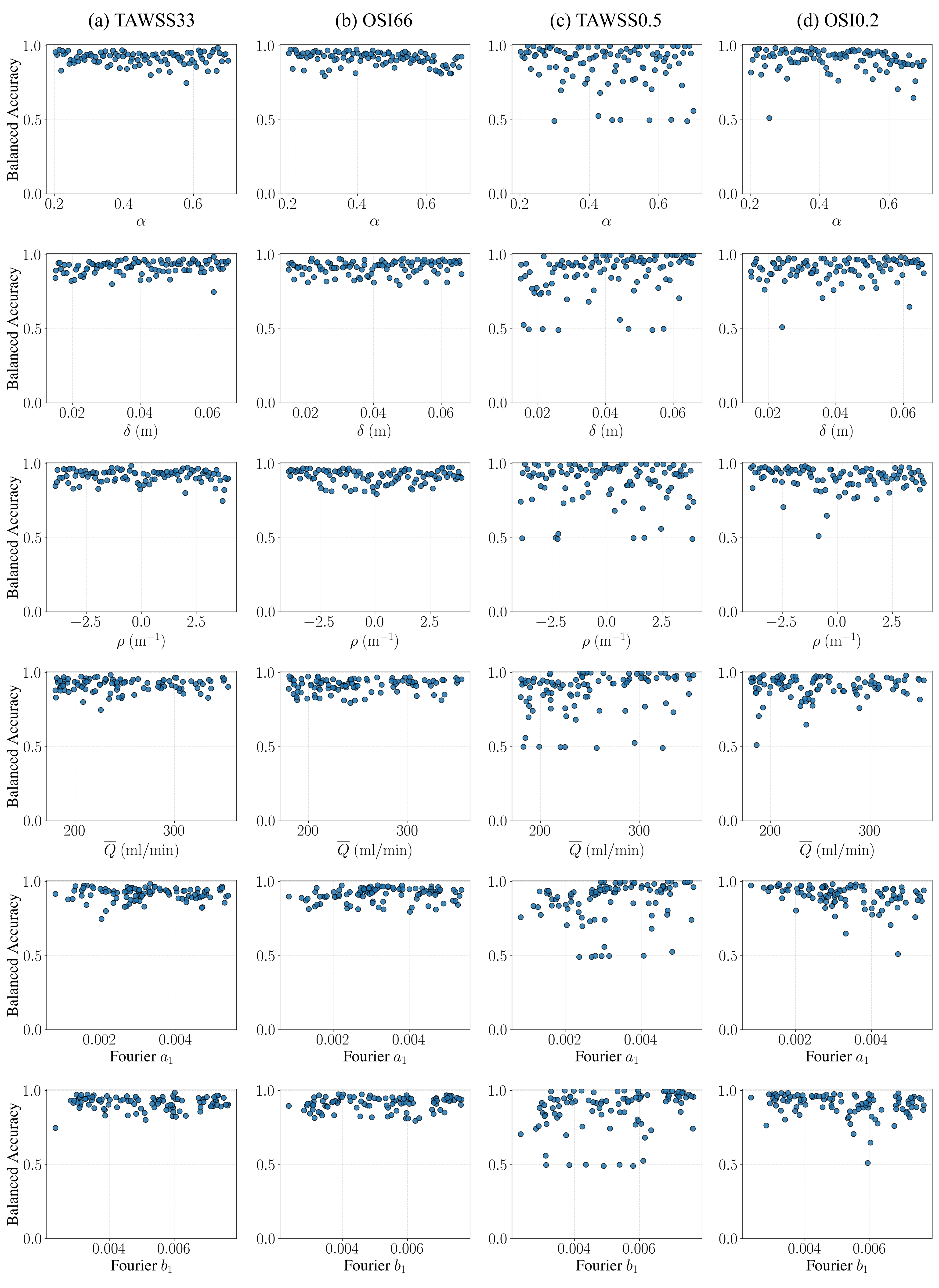}
\caption{BA of the Fourier models across the test dataset plotted against selected geometric and waveform parameters for the four critical-region definitions: (a) TAWSS33, (b) OSI66, (c) TAWSS0.5, and (d) OSI0.2. Results for other parameters are reported in Supplementary Material 2.}
\label{fig:BA_vs_some_inputs}
\end{figure}

The robustness of the Fourier models were examined by plotting BA against selected input parameters, namely stenosis severity ($\alpha$), stenosis length ($\delta$), curvature ($\rho$), average flowrate ($\bar{Q}$) and the first Fourier coefficient pair ($a_1$ and $b_1$), as in Figure \ref{fig:BA_vs_some_inputs}. Across all four critical-region definitions, no clear monotonic trends were observed between model performance and individual input features. BA remained consistently high over the full range of parameter values, indicating that the model generalises effectively across diverse anatomical and waveform conditions rather than being tuned to narrow subsets of the input space. Mild differences can be observed between thresholding approaches: critical regions with relative thresholds (TAWSS33 in Figure \ref{fig:BA_vs_some_inputs}a and OSI66 in Figure \ref{fig:BA_vs_some_inputs}b) yielded slightly higher and more consistent BA compared to absolute thresholds (TAWSS0.5 in Figure \ref{fig:BA_vs_some_inputs}c and OSI0.2 in Figure \ref{fig:BA_vs_some_inputs}d). This suggests that percentile-based thresholds produce critical regions of more consistent size and distribution across cases, whereas absolute thresholds can generate more fragmented or imbalanced regions, making them harder to predict reliably. Similar BA plots against the rest of the input features are provided in Supplementary Material 2.

\begin{figure}
\centering
\includegraphics[width=0.7\textwidth]{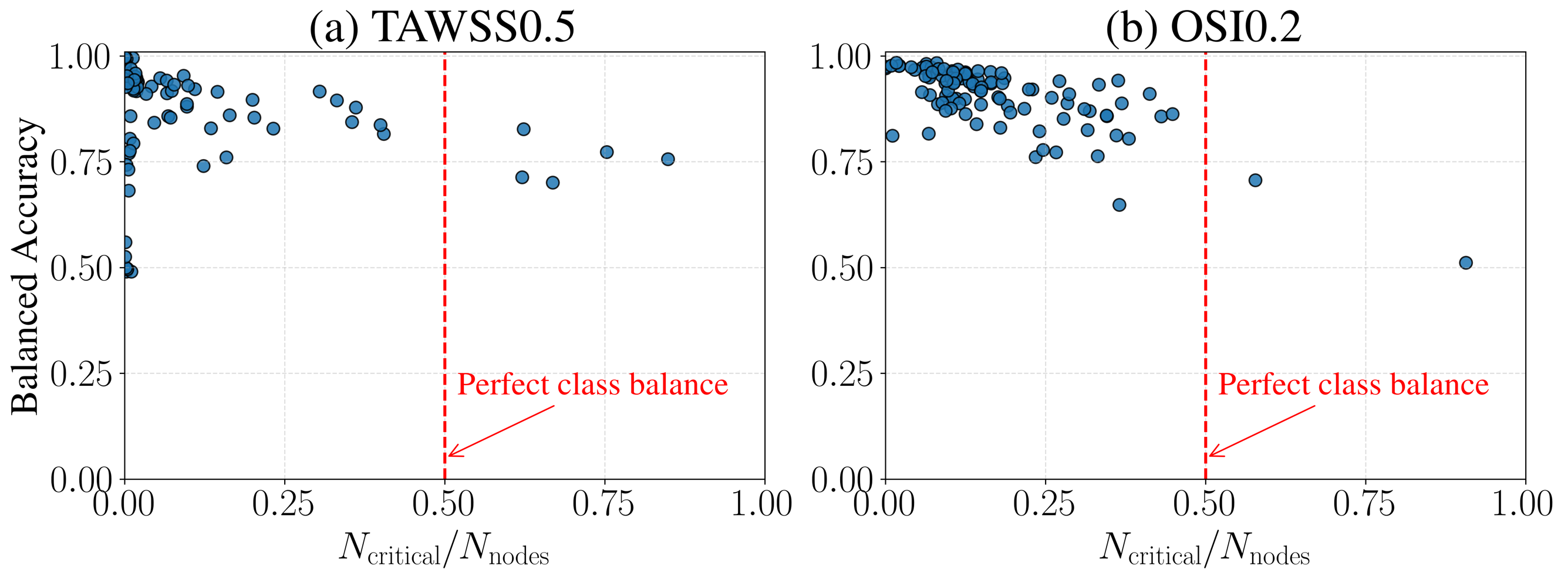}
\caption{
BA of Fourier models as a function of class imbalance, quantified by the fraction of nodes classified as critical ($N_{\mathrm{critical}}/N_{\mathrm{nodes}}$) for (a) TAWSS0.5 and (b) OSI0.2. The vertical dashed line marks the point of perfect class balance.}
\label{fig:BA_vs_class_imbalance}
\end{figure}

Lastly, the relationship between model performance and class imbalance was examined by plotting BA against the fraction of nodes classified as critical ($N_{\mathrm{critical}}/N_{\mathrm{nodes}}$) in Figure \ref{fig:BA_vs_class_imbalance}. This analysis was performed only for TAWSS0.5 and OSI0.2, since by definition, TAWSS33 and OSI66 always include approximately $33\%$ of nodes in the critical regions, and thus do not vary in class balance. TAWSS0.5 (Figure \ref{fig:BA_vs_class_imbalance}a) and OSI0.2 (Figure \ref{fig:BA_vs_class_imbalance}b) produced highly skewed distributions, with most cases concentrated near the lower end of the critical-node fraction axis because absolute thresholds often identify only a small proportion of the vessel surface as part of the critical regions. A mild negative correlation between BA and critical-node fraction was observed, which is likely because cases with low critical nodes dominated the training data. A number of TAWSS0.5 cases achieved a perfect BA = 1.00 because no nodes were classified as critical, and the model correctly captured this outcome\footnote{In such cases, the calculation of BA in Equation \ref{eq:BA} involve 0/0, which we then corrected to 1.00 to represent the perfectly correct prediction.}. The vertical red line at critical-node fraction of 0.5 highlights the point of perfect class balance, which was rarely approached in either index. These findings indicate that class imbalance is intrinsic to absolute thresholding, and this contributes to the greater variability observed in predictive performance compared with relative thresholds.

\subsection{Computational cost}

\begin{table}
\centering
\caption{Offline and online computational costs of the Fourier models for the four critical-region definitions: TAWSS33, OSI66, TAWSS0.5, and OSI0.2. $\mu$s = microseconds.}
\label{tab:comp_cost}
\begin{tabular}{llcccc}
\hline
                            &                               & \textbf{TAWSS33} & \textbf{OSI66} & \textbf{TAWSS0.5} & \textbf{OSI0.2}      \\ \hline
\multicolumn{1}{c}{Offline} & Dataset generation time (CFD) & 600 hours        & 600 hours      & 600 hours         & 600 hours            \\
\multicolumn{1}{c}{}        & POD ROM construction time     & 0.295 s          & 0.256 s        & 0.284 s           & 0.260 s              \\
\multicolumn{1}{c}{}        & ML development time           & 249 mins         & 289 mins       & 323 mins          & 320 mins             \\ \hline
\multicolumn{1}{c}{Online}  & CFD simulation time           & 40 hours         & 40 hours       & 40 hours          & 40 hours             \\
                            & ML evaluation time            & 74.7 $\mu$s      & 83.5 $\mu$s    & 94.3 $\mu$s       & 79.3 $\mu$s           \\
                            & Speed-up ratio                & $1.93\times10^9 $ & $1.72\times10^9 $ & $1.53\times10^9 $ & $1.82\times10^9 $ \\ \hline
                            & \multicolumn{1}{c}{}          &                  &                &                   & \multicolumn{1}{l}{} \\ 
                            & \multicolumn{1}{c}{}          &                  &                &                   & \multicolumn{1}{l}{}
\end{tabular}
\end{table}

A major advantage of the proposed framework lies in its substantial reduction of computational time cost, shown in Table \ref{tab:comp_cost}. The full-order CFD simulations were performed on the UCL Department of Computer Science cluster (Intel Xeon Gold 5118 at 2.3 GHz, 10 processors). To generate the dataset of 500 cases, simulations were divided into 10 batches. Depending on case setup and available computing resources in the cluster, each batch required 20–60 hours to run. Following completion, the CFD results from each batch were compressed and transferred to the local workstation, adding an additional 20–25 hours. Certain stages overlapped in execution: for example, new CFD runs were initiated as soon as the previous batch had been compressed and prepared for download. Occasional cluster failures interrupted some CFD cases and required reruns, which prolonged the process. In total, dataset generation consumed approximately 25 days (or 600 hours). Four cases continued to fail at the final rerun stage and were excluded, leaving 469 cases for downstream analysis\footnote{Thus, the actual training:validation:test split is 300:97:99.}.

The ROM and ML tasks were performed locally. POD construction for each critical-region definition was executed on a local workstation CPU (Intel(R) Core(TM) i9-12900K at 3.2 GHz), requiring less than 0.30 seconds, highlighting the negligible cost of building the ROMs once the CFD dataset is available. The ML development stage consisted of an Optuna-based hyperparameter search performed on the local CPU, where each Optuna trial launched 10 consecutive training runs on an Nvidia RTX A2000 GPU. Across targets, the ML development time in each critical-region definition ranges from 249 to 323 minutes.

Online evaluation of full trained ML models on new inputs is effectively instantaneous. Inference was performed on the local CPU, with each evaluation taking less than 100 microseconds, compared to approximately 40 hours for a single CFD simulation. This corresponds to speed-up ratios on the order of $10^9$. It is important to emphasise that the reported speed-up ratios are estimates, as the CFD simulations and the ML evaluation were executed on different computational platforms.

\section{Discussion}

The results demonstrate that the proposed ML-ROM framework can accurately identify critical regions defined by both artery-specific WSS percentiles (TAWSS33, OSI66) and fixed absolute cut-offs (TAWSS0.5, OSI0.2). The Fourier representation of inflow waveforms proved most effective, outperforming LSTM and CNN. This suggests that a compact Fourier representation of the inflow waveform was sufficient to achieve a reliable prediction of critical regions. Excessive temporal information may act as noise, and thus reduce the predictive performance of ML models. Specifically, only a few dominant frequency contents carry the most useful predictive value, as revealed by the Optuna hyperparameter fine-tuning results: only 1-3 Fourier pairs were needed. However, this likely stems from the decision to simplify the task to predicting binary critical-region fields rather than full TAWSS or OSI fields, which reduces dependence on detailed flowrate waveform information. In fact, prior studies that predicted full haemodynamic fields under varying flowrate waveforms in fixed geometries have shown that the predicted fields change with variations in the input waveform. Examples include the prediction of WSS by Chatpattanasiri et al. \cite{chatpattanasiri2025ML}, as well as velocity fields by MacRaild et al. \cite{macraild2024reduced} and Drakoulas et al. \cite{drakoulas2023fastsvd}. In this context, compact Fourier representations provide effective inputs for the current critical-region prediction framework, while greater temporal resolution may be needed when the objective is to reproduce full haemodynamic quantities rather than thresholded critical regions.

Robustness analyses confirmed that the prediction performance of the Fourier models was unaffected by variability in geometry and inflow waveform. The models achieved excellent generalisability across cases inside the test dataset. Predictions based on TAWSS0.5 and OSI0.2 showed slightly lower performance compared to those based on TAWSS33 and OSI66, driven by class imbalance. This is consistent with the POD reconstruction analysis, where TAWSS33 and OSI66-based ROMs yielded more stable and accurate reconstructions than TAWSS0.5 and OSI0.2 ones. Findings from both POD and ML stages lead to a similar conclusion: percentile-based thresholding definitions produce more consistent critical regions, whereas absolute thresholding definitions may create imbalanced datasets that are inherently harder to model \cite{altalhan2025imbalanced, banerjee2023machine}. 






An important limitation of the present work concerns the type of output predicted. In this study, the framework targeted binary fields of critical regions. This choice was made to keep computational demand tractable and facilitate model training. While effective for highlighting risk-prone areas, this inevitably discards richer haemodynamic detail. Predicting continuous TAWSS and OSI distributions, or more ambitiously, the full time-dependent WSS distributions, would provide much greater physiological insight. However, achieving this while taking into account both geometric and waveform variability would likely require more advanced dimensionality reduction methods than POD, which is a fully linear method \cite{Brunton_Kutz_2022, Berkooz1993ThePOD}. 

An alternative would be to bypass dimensionality reduction altogether by predicting WSS point-wise in spatio-temporal coordinates. This might necessitate more advanced model architectures. For example, graph convolutional networks (GCNs) can naturally capture spatial dependencies on irregular vascular meshes \cite{bhatti2023deep, yao2024image2flow, pegolotti2024learning}. Notable examples include Image2Flow by Yao et al. \cite{yao2024image2flow}, which used a U-net architecture with the graph structure of GCNs to represent vascular mesh topology. Their model performed segmentation, mesh generation, and haemodynamic field prediction in a single framework. Another interesting model architecture is hypernetworks, where one network generates the weights of another, offering a means to adapt parameters across varying geometrical and physiological conditions and thereby improving generalisability. \cite{pan2023NIF, chauhan2024brief}. The Neural Implicit Flow (NIF) by Pan et al. \cite{pan2023NIF} framework illustrates how hypernetworks can be applied, with the main network predicting field values from the spatial coordinate and the hypernetwork taking flow conditions (e.g. time, Reynolds number, sensor measurements) as the input to generate parameters in the main network. With these more advance network architectures, vessel geometries can be incorporated into the network in greater detail. Rather than relying on a limited set of geometric parameters, the input could include full vessel spatial coordinates (with mesh connectivity for GCNs). Following this direction, synthetic geometries may be generated using more complex methods such as statistical shape models (SSMs) or generative adversarial networks (GANs), which capture more detailed anatomical variability to generate more realistic vascular geometries. Examples include the work of Pajaziti et al. \cite{pajaziti2023shape} who generated 3000 synthetic aortic shapes from the original 67 geometries using SSM. These approaches could enable richer training datasets and would lead to better prediction accuracy and generalisability.

Finally, the framework was developed as a purely data-driven approach, with no explicit incorporation of physical governing equations. Embedding physics-informed elements, for instance through physics-informed neural networks (PINNs) \cite{cai2021physics}, may improve predictive accuracy and interpretability. For example, Csala et al. \cite{csala2025physics} developed a physics-constrained coupled neural differential equation framework that improves the accuracy of one-dimensional cardiovascular flow models. More similar to our current work, Wong et al. \cite{wong2024strategies} employed PINNs to predict velocity and pressure fields across multiple two-dimensional idealised stenosed cylindrical geometries, parameterised by stenosis length and severity. By implementing PINNs within a Hypernetwork framework, Wong et al. \cite{wong2024strategies} achieved notable accuracy and robustness. This performance, however, came at the expense of extremely long runtimes and substantial GPU demand. Chan et al. \cite{chan2025role} extended this idea into three-dimensional idealised stenosed tube-like geometries with sinusoidal shape curvature. Interestingly, they found that fully data-driven models outperform PINNs, in both accuracy and training cost. This can be attributed to the large and quality training dataset, arranged nicely in a uniform grid in the geometric parameter space. By contrast, in settings like ours where assembling such abundant, uniformly distributed data is impractical, integrating PINNs into the ML-ROM framework may help improve the model generalisability. The key challenge lies in designing an efficient coupling between PINNs and ROM. Past attempts to incorporate PINNs with POD-based ROM include the works of Chen et al. \cite{chen2021physics} and Fu et al \cite{fu2023physics}, but their works remained confined to basic case studies with simple idealised geometries. These studies highlight both the promise and the practical challenges of merging PINNs with ROM, particularly when moving beyond simplified cases.

\section{Conclusions}
This work presents a novel ML–ROM framework to identify restenosis-prone regions on the surface of femoral arteries. To the best of the author's knowledge, our study is the first to incorporate variability in vessel geometry and inflow waveform simultaneously in prediction tasks. Synthetic femoral artery geometries and realistic inlet flowrate waveforms were generated, with CFD providing WSS distributions from which critical regions were identified using both relative (TAWSS33, OSI66) and absolute (TAWSS0.5, OSI0.2) thresholds. POD was applied to obtain compact reduced-order representations of the critical-region distributions, and three ML model architectures (Fourier, LSTM, and CNN models) were trained to predict the corresponding POD coefficients, which were subsequently used to reconstruct the predicted critical regions. The Fourier models emerged as the most effective, achieving test-set BA above 0.92 across all critical-region definitions, and demonstrating strong robustness to diverse anatomical and waveform conditions. A small number of Fourier coefficients efficiently captures inflow waveform dynamics that are essential to predict the critical-region distributions, outperforming more complex LSTM and CNN models that are trained on the full waveform sequence. The framework delivers fast critical-region identification, with a speed-up ratio on the order of $10^9$ compared to the time it takes to run a case of CFD. By providing accurate and generalisable identification of restenosis-prone regions under realistic variability, this study offers a powerful and computationally efficient route towards personalised vascular risk assessment and paves the way for ML-ROM methodologies into clinical decision support.

\section*{CRediT authorship contribution statement}

\textbf{Chotirawee Chatpattanasiri}: Writing - original draft, Conceptualization, Visualization, Validation, Methodology, Investigation, Formal analysis. 
\textbf{Federica Ninno}: Software, Data Curation, Validation, Investigation, Formal analysis.
\textbf{Vanessa D\'{\i}az-Zuccarini}: Writing - review \& editing, Supervision, Resources, Project administration, Funding acquisition, Conceptualization.
\textbf{Stavroula Balabani}: Writing - review \& editing, Supervision, Resources, Project administration, Funding acquisition, Conceptualization.


\section*{Declaration of competing interest}
The authors declare that they have no known competing financial interests or personal relationships that could have appeared to influence the work reported in this paper.

\section*{Acknowledgments}
This project has been supported by 
UK Research and Innovation (UKRI) (BB/X005062/1);
British Heart Foundation (NH/20/1/34705);
the Biotechnology and Biological Sciences Research Council (BBSRC);
University College London EPSRC Centre for Doctoral Training i4health (EP/S021930/1);
the EPSRC Research Grant ``Hidden haemodynamics: A Physics-InfOrmed, real-time recoNstruction framEwork for haEmodynamic virtual pRototyping and clinical support (PIONEER)" (EP/W00481X/1)
UCL Centre for Digital Innovation (CDI) powered by Amazon Web Service (AWS);
and the Department of Mechanical Engineering, University College London.
The authors also thank UCL Department of Computer Science for providing high-performance computing resources.

\clearpage







\clearpage

\printbibliography

@article{fowkes2017peripheral,
  title={Peripheral artery disease: epidemiology and global perspectives},
  author={Fowkes, F Gerry R and Aboyans, Victor and Fowkes, Freya JI and McDermott, Mary M and Sampson, Uchechukwu KA and Criqui, Michael H},
  journal={Nature Reviews Cardiology},
  volume={14},
  number={3},
  pages={156--170},
  year={2017},
  publisher={Nature Publishing Group UK London},
  doi={10.1038/nrcardio.2016.179}
}

@article{SHAMAKI2022101082,
title = {Peripheral Artery Disease: A Comprehensive Updated Review},
journal = {Current Problems in Cardiology},
volume = {47},
number = {11},
pages = {101082},
year = {2022},
issn = {0146-2806},
doi = {10.1016/j.cpcardiol.2021.101082},
url = {https://www.sciencedirect.com/science/article/pii/S0146280621002905},
author = {Garba Rimamskep Shamaki and Favour Markson and Demilade Soji-Ayoade and Chibuike Charles Agwuegbo and Michael Olaseni Bamgbose and Bob-Manuel Tamunoinemi}
}

@article{Houghton2024,
    author = {Houghton, John S M and Saratzis, Athanasios N and Sayers, Rob D and Haunton, Victoria J},
    title = {New Horizons in Peripheral Artery Disease},
    journal = {Age and Ageing},
    volume = {53},
    number = {6},
    pages = {afae114},
    year = {2024},
    month = {06},
    issn = {1468-2834},
    doi = {10.1093/ageing/afae114},
    url = {https://doi.org/10.1093/ageing/afae114},
    eprint = {https://academic.oup.com/ageing/article-pdf/53/6/afae114/58238308/afae114.pdf},
}

@article{colombo2021baseline,
  title={{Baseline local hemodynamics as predictor of lumen remodeling at 1-year follow-up in stented superficial femoral arteries}},
  author={Colombo, Monika and He, Yong and Corti, Anna and Gallo, Diego and Casarin, Stefano and Rozowsky, Jared M and Migliavacca, Francesco and Berceli, Scott and Chiastra, Claudio},
  journal={Scientific Reports},
  volume={11},
  number={1},
  pages={1613},
  year={2021},
  publisher={Nature Publishing Group UK London},
  doi={10.1038/s41598-020-80681-8}
}

@article{donadoni2020multiscale,
  title={Multiscale, patient-specific computational fluid dynamics models predict formation of neointimal hyperplasia in saphenous vein grafts},
  author={Donadoni, Francesca and Pichardo-Almarza, Cesar and Homer-Vanniasinkam, Shervanthi and Dardik, Alan and D{\'\i}az-Zuccarini, Vanessa},
  journal={Journal of Vascular Surgery Cases, Innovations and Techniques},
  volume={6},
  number={2},
  pages={292--306},
  year={2020},
  publisher={Elsevier},
  doi={10.1016/j.jvscit.2019.09.009}
}

@article{Ninno2024modelling_lower_limb,
  title = {{Modelling lower-limb peripheral arterial disease using clinically available datasets: impact of inflow boundary conditions on hemodynamic indices for restenosis prediction}},
  journal = {Computer Methods and Programs in Biomedicine},
  pages = {108214},
  year = {2024},
  issn = {0169-2607},
  doi = {https://doi.org/10.1016/j.cmpb.2024.108214},
  url = {https://www.sciencedirect.com/science/article/pii/S0169260724002062},
  author = {Federica Ninno and Claudio Chiastra and Monika Colombo and Alan Dardik and David Strosberg and Edouard Aboian and Janice Tsui and Matthew Bartlett and Stavroula Balabani and Vanessa Díaz-Zuccarini},
}

@article{ninno2023systematic,
  title={{A systematic review of clinical and biomechanical engineering perspectives on the prediction of restenosis in coronary and peripheral arteries}},
  author={Ninno, Federica and Tsui, Janice and Balabani, Stavroula and D{\'\i}az-Zuccarini, Vanessa},
  journal={JVS-Vascular Science},
  volume={4},
  pages={100128},
  year={2023},
  publisher={Elsevier},
  doi={10.1016/j.jvssci.2023.100128}
}

@article{colombo2020computing,
  title={{Computing patient-specific hemodynamics in stented femoral artery models obtained from computed tomography using a validated 3D reconstruction method}},
  author={Colombo, Monika and Bologna, Marco and Garbey, Marc and Berceli, Scott and He, Yong and Matas, Jos{\`e} Felix Rodriguez and Migliavacca, Francesco and Chiastra, Claudio},
  journal={Medical engineering \& physics},
  volume={75},
  pages={23--35},
  year={2020},
  publisher={Elsevier},
  doi={10.1016/j.medengphy.2019.10.005}
}

@article{colombo2022superficial,
  title={{Superficial femoral artery stenting: Impact of stent design and overlapping on the local hemodynamics}},
  author={Colombo, Monika and Corti, Anna and Gallo, Diego and Colombo, Andrea and Antognoli, Giacomo and Bernini, Martina and McKenna, Ciara and Berceli, Scott and Vaughan, Ted and Migliavacca, Francesco and others},
  journal={Computers in Biology and Medicine},
  volume={143},
  pages={105248},
  year={2022},
  publisher={Elsevier},
  doi={10.1016/j.compbiomed.2022.105248}
}

@article{bonfanti2017computational,
  title={Computational tools for clinical support: a multi-scale compliant model for haemodynamic simulations in an aortic dissection based on multi-modal imaging data},
  author={Bonfanti, Mirko and Balabani, Stavroula and Greenwood, John P and Puppala, Sapna and Homer-Vanniasinkam, Shervanthi and D{\'\i}az-Zuccarini, Vanessa},
  journal={Journal of The Royal Society Interface},
  volume={14},
  number={136},
  pages={20170632},
  year={2017},
  publisher={The Royal Society},
  doi={10.1098/rsif.2017.0632}
}

@article{Bonfanti2020,
   author = {Mirko Bonfanti and Gaia Franzetti and Shervanthi Homer-Vanniasinkam and Vanessa Díaz-Zuccarini and Stavroula Balabani},
   doi = {10.1007/s10439-020-02603-z},
   issn = {15739686},
   issue = {12},
   journal = {Annals of Biomedical Engineering},
   month = {12},
   pages = {2950-2964},
   pmid = {32929558},
   publisher = {Springer},
   title = {{A Combined In Vivo, In Vitro, In Silico Approach for Patient-Specific Haemodynamic Studies of Aortic Dissection}},
   volume = {48},
   year = {2020},
}

@article{Stokes2021,
   author = {Catriona Stokes and Mirko Bonfanti and Zeyan Li and Jiang Xiong and Duanduan Chen and Stavroula Balabani and Vanessa Díaz-Zuccarini},
   doi = {10.1016/j.jbiomech.2021.110793},
   issn = {18732380},
   journal = {Journal of Biomechanics},
   month = {12},
   pmid = {34715606},
   publisher = {Elsevier Ltd},
   title = {{A novel MRI-based data fusion methodology for efficient, personalised, compliant simulations of aortic haemodynamics}},
   volume = {129},
   year = {2021},
}

@article{Stokes2023The_Influence,
   author = {C. Stokes and F. Haupt and D. Becker and V. Muthurangu and H. von Tengg-Kobligk and S. Balabani and V. Díaz-Zuccarini},
   doi = {10.1007/s10439-023-03175-4},
   issn = {15739686},
   journal = {Annals of Biomedical Engineering},
   publisher = {Springer},
   title = {{The Influence of Minor Aortic Branches in Patient-Specific Flow Simulations of Type-B Aortic Dissection}},
   year = {2023},
}

@article{stokes2023aneurysmal,
author={Stokes, C and Ahmed, D and Lind, N and Haupt, Fabian and Becker, Daniel and Hamilton, J and Muthurangu, V and von Tengg-Kobligk, Hendrik and Papadakis, G and Balabani, S and others},
doi = {10.1098/rsif.2023.0281},
issn = {17425662},
issue = {206},
journal = {Journal of the Royal Society Interface},
month = {9},
pmid = {37727072},
publisher = {Royal Society Publishing},
title = {{Aneurysmal growth in type-B aortic dissection: assessing the impact of patient-specific inlet conditions on key haemodynamic indices}},
volume = {20},
year = {2023},
}

@article{candreva2022current,
  title={Current and future applications of computational fluid dynamics in coronary artery disease},
  author={Candreva, Alessandro and De Nisco, Giuseppe and Rizzini, Maurizio Lodi and D’Ascenzo, Fabrizio and De Ferrari, Gaetano Maria and Gallo, Diego and Morbiducci, Umberto and Chiastra, Claudio},
  journal={Reviews in Cardiovascular Medicine},
  volume={23},
  number={11},
  pages={377},
  year={2022},
  publisher={IMR press},
  doi={10.31083/j.rcm2311377}
}

@article{psiuk2024methodology,
  title={{Methodology of generation of CFD meshes and 4D shape reconstruction of coronary arteries from patient-specific dynamic CT}},
  author={Psiuk-Maksymowicz, Krzysztof and Borys, Damian and Melka, Bartlomiej and Gracka, Maria and Adamczyk, Wojciech P and Rojczyk, Marek and Wasilewski, Jaroslaw and G{\l}owacki, Jan and Kruk, Mariusz and Nowak, Marcin and others},
  journal={Scientific Reports},
  volume={14},
  number={1},
  pages={2201},
  year={2024},
  publisher={Nature Publishing Group UK London},
  doi={10.1038/s41598-024-52398-5}
}

@article{carbonaro2023design,
  title={Design of innovative self-expandable femoral stents using inverse homogenization topology optimization},
  author={Carbonaro, Dario and Mezzadri, Francesco and Ferro, Nicola and De Nisco, Giuseppe and Audenino, Alberto Luigi and Gallo, Diego and Chiastra, Claudio and Morbiducci, Umberto and Perotto, Simona},
  journal={Computer Methods in Applied Mechanics and Engineering},
  volume={416},
  pages={116288},
  year={2023},
  publisher={Elsevier},
  doi={10.1016/j.cma.2023.116288}
}

@article{corti2024impact,
  title={Impact of tissue damage and hemodynamics on restenosis following percutaneous transluminal angioplasty: a patient-specific multiscale model},
  author={Corti, Anna and Marradi, Matilde and {\c{C}}elikbudak Orhon, Cemre and Boccafoschi, Francesca and B{\"u}chler, Philippe and Rodriguez Matas, Jose F and Chiastra, Claudio},
  journal={Annals of Biomedical Engineering},
  volume={52},
  number={8},
  pages={2203--2220},
  year={2024},
  publisher={Springer},
  doi={10.1007/s10439-024-03520-1}
}

@article{gokgol2019prediction,
  title={Prediction of restenosis based on hemodynamical markers in revascularized femoro-popliteal arteries during leg flexion},
  author={G{\"o}kg{\"o}l, Can and Diehm, Nicolas and R{\"a}ber, Lorenz and B{\"u}chler, Philippe},
  journal={Biomechanics and modeling in mechanobiology},
  volume={18},
  number={6},
  pages={1883--1893},
  year={2019},
  publisher={Springer},
  doi={10.1007/s10237-019-01183-9}
}

@article{hoogendoorn2020multidirectional,
    title={Multidirectional wall shear stress promotes advanced coronary plaque development: comparing five shear stress metrics},
    author={Hoogendoorn, Ayla and Kok, Annette M and Hartman, Eline MJ and de Nisco, Giuseppe and Casadonte, Lorena and Chiastra, Claudio and Coenen, Adriaan and Korteland, Suze-Anne and Van der Heiden, Kim and Gijsen, Frank JH and others},
    journal={Cardiovascular research},
    volume={116},
    number={6},
    pages={1136--1146},
    year={2020},
    publisher={Oxford University Press},
    doi={10.1093/cvr/cvz212}
}

@article{de2024predicting,
  title={Predicting lipid-rich plaque progression in coronary arteries using multimodal imaging and wall shear stress signatures},
  author={De Nisco, Giuseppe and Hartman, Eline MJ and Torta, Elena and Daemen, Joost and Chiastra, Claudio and Gallo, Diego and Morbiducci, Umberto and Wentzel, Jolanda J},
  journal={Arteriosclerosis, Thrombosis, and Vascular Biology},
  volume={44},
  number={4},
  pages={976--986},
  year={2024},
  publisher={Lippincott Williams \& Wilkins Hagerstown, MD},
  doi={10.1161/ATVBAHA.123.320337}
}

@article{BONFANTI2018,
title = {{A simplified method to account for wall motion in patient-specific blood flow simulations of aortic dissection: Comparison with fluid-structure interaction}},
journal = {Medical Engineering \& Physics},
volume = {58},
pages = {72-79},
year = {2018},
issn = {1350-4533},
doi = {https://doi.org/10.1016/j.medengphy.2018.04.014},
url = {https://www.sciencedirect.com/science/article/pii/S1350453318300742},
author = {Mirko Bonfanti and Stavroula Balabani and Mona Alimohammadi and Obiekezie Agu and Shervanthi Homer-Vanniasinkam and Vanessa Díaz-Zuccarini},
}

@article{chatpattanasiri2025ML,
    doi = {10.1371/journal.pone.0325644},
    author = {Chatpattanasiri, Chotirawee AND Ninno, Federica AND Stokes, Catriona AND Dardik, Alan AND Strosberg, David AND Aboian, Edouard AND von Tengg-Kobligk, Hendrik AND Díaz-Zuccarini, Vanessa AND Balabani, Stavroula},
    journal = {PLOS ONE},
    publisher = {Public Library of Science},
    title = {ML-ROM wall shear stress prediction in patient-specific vascular pathologies under a limited clinical training data regime},
    year = {2025},
    month = {06},
    volume = {20},
    url = {https://doi.org/10.1371/journal.pone.0325644},
    pages = {1-25},
    number = {6},
}

@article{Liang2019A,
title={{A feasibility study of deep learning for predicting hemodynamics of human thoracic aorta}},
author={L. Liang and W. Mao and Wei Sun},
journal={Journal of biomechanics},
year={2019},
pages={ 109544 },
doi={10.1016/j.jbiomech.2019.109544}
}

@article{pajaziti2023shape,
  title={{Shape-driven deep neural networks for fast acquisition of aortic 3D pressure and velocity flow fields}},
  author={Pajaziti, Endrit and Montalt-Tordera, Javier and Capelli, Claudio and Sivera, Rapha{\"e}l and Sauvage, Emilie and Quail, Michael and Schievano, Silvia and Muthurangu, Vivek},
  journal={PLoS Computational Biology},
  volume={19},
  number={4},
  pages={e1011055},
  year={2023},
  publisher={Public Library of Science San Francisco, CA USA},
  doi={10.1371/journal.pcbi.1011055}
}

@article{siena2023data,
  title={{Data-driven reduced order modelling for patient-specific hemodynamics of coronary artery bypass grafts with physical and geometrical parameters}},
  author={Siena, Pierfrancesco and Girfoglio, Michele and Ballarin, Francesco and Rozza, Gianluigi},
  journal={Journal of Scientific Computing},
  volume={94},
  number={2},
  pages={38},
  year={2023},
  publisher={Springer},
  doi={10.1007/s10915-022-02082-5}
}

@article{drakoulas2023fastsvd,
  title={{FastSVD-ML--ROM: A reduced-order modeling framework based on machine learning for real-time applications}},
  author={Drakoulas, GI and Gortsas, Theodore V and Bourantas, George C and Burganos, Vasilis N and Polyzos, Demosthenes},
  journal={Computer Methods in Applied Mechanics and Engineering},
  volume={414},
  pages={116155},
  year={2023},
  publisher={Elsevier},
  doi={10.1016/j.cma.2023.116155}
}

@article{alamir2024rapid,
  title={Rapid prediction of wall shear stress in stenosed coronary arteries based on deep learning},
  author={Alamir, Salwa Husam and Tufaro, Vincenzo and Trilli, Matilde and Kitslaar, Pieter and Mathur, Anthony and Baumbach, Andreas and Jacob, Joseph and Bourantas, Christos V and Torii, Ryo},
  journal={Frontiers in Bioengineering and Biotechnology},
  volume={12},
  pages={1360330},
  year={2024},
  publisher={Frontiers Media SA},
  doi={10.3389/fbioe.2024.1360330}
}

@article{Ferdian20204DFlowNet,
title={{4DFlowNet: Super-Resolution 4D Flow MRI Using Deep Learning and Computational Fluid Dynamics}},
author={Ferdian, Edward and Suinesiaputra, Avan and Dubowitz, David J and Zhao, Debbie and Wang, Alan and Cowan, Brett and Young, Alistair A},
journal={Frontiers in Physics},
volume={8},
pages={138},
year={2020},
publisher={Frontiers Media SA},
doi={10.3389/fphy.2020.00138}
}

@article{fathi2020super,
  title={{Super-resolution and denoising of 4D-flow MRI using physics-informed deep neural nets}},
  author={Fathi, Mojtaba F and Perez-Raya, Isaac and Baghaie, Ahmadreza and Berg, Philipp and Janiga, Gabor and Arzani, Amirhossein and D’Souza, Roshan M},
  journal={Computer Methods and Programs in Biomedicine},
  volume={197},
  pages={105729},
  year={2020},
  publisher={Elsevier},
  doi={10.1016/j.cmpb.2020.105729}
}

@article{kontogiannis2022joint,
  title={Joint reconstruction and segmentation of noisy velocity images as an inverse Navier--Stokes problem},
  author={Kontogiannis, Alexandros and Elgersma, Scott V and Sederman, Andrew J and Juniper, Matthew P},
  journal={Journal of Fluid Mechanics},
  volume={944},
  pages={A40},
  year={2022},
  publisher={Cambridge University Press},
  doi={10.1017/jfm.2022.503}
}

@article{Shlezinger2020Model,
title={{Model-Based Deep Learning}},
author={Nir Shlezinger and Jay Whang and Yonina C. Eldar and A. Dimakis},
journal={Proceedings of the IEEE},
year={2020},
volume={111},
pages={465-499},
doi={10.1109/JPROC.2023.3247480}
}

@book{alpaydin2020introduction_to_ML,
  title={{Introduction to machine learning}},
  author={Alpaydin, Ethem},
  year={2020},
  publisher={MIT press},
  ISBN={978-0-262-01243-0}
}

@article{Arzani2021,
   author = {Amirhossein Arzani and Scott T.M. Dawson},
   doi = {10.1098/rsif.2020.0802},
   issn = {17425662},
   issue = {175},
   journal = {Journal of the Royal Society Interface},
   month = {2},
   pmid = {33561376},
   publisher = {Royal Society Publishing},
   title = {{Data-driven cardiovascular flow modelling: Examples and opportunities}},
   volume = {18},
   year = {2021},
}

@book{Brunton_Kutz_2022, 
place={Cambridge}, 
edition={2}, 
title={{Data-Driven Science and Engineering: Machine Learning, Dynamical Systems, and Control}}, 
publisher={Cambridge University Press}, 
author={Brunton, Steven L. and Kutz, J. Nathan}, 
year={2022}
}

@article{Du2018Dimensionality,
title={{Dimensionality Reduction Techniques for Visualizing Morphometric Data: Comparing Principal Component Analysis to Nonlinear Methods}},
author={Trina Y. Du},
journal={Evolutionary Biology},
year={2018},
volume={46},
pages={106 - 121},
doi={10.1007/s11692-018-9464-9}
}

@article{di2019reduced,
  title={{Reduced-order modeling of left ventricular flow subject to aortic valve regurgitation}},
  author={Di Labbio, Giuseppe and Kadem, Lyes},
  journal={Physics of Fluids},
  volume={31},
  number={3},
  year={2019},
  publisher={AIP Publishing},
  doi={10.1063/1.5083054}
}

@article{csala2024comparison,
  title={A comparison of machine learning methods for recovering noisy and missing 4D flow MRI data},
  author={Csala, Hunor and Amili, Omid and D'Souza, Roshan M and Arzani, Amirhossein},
  journal={International Journal for Numerical Methods in Biomedical Engineering},
  volume={40},
  number={11},
  pages={e3858},
  year={2024},
  publisher={Wiley Online Library},
  doi={https://doi.org/10.1002/cnm.3858}
}

@article{macraild2024reduced,
  title={Reduced order modelling of intracranial aneurysm flow using proper orthogonal decomposition and neural networks},
  author={MacRaild, Michael and Sarrami-Foroushani, Ali and Lassila, Toni and Frangi, Alejandro F},
  journal={International Journal for Numerical Methods in Biomedical Engineering},
  volume={40},
  number={10},
  pages={e3848},
  year={2024},
  publisher={Wiley Online Library},
  doi={10.1002/cnm.3848}
}

@article{barzegar2025predictive,
  title={A predictive surrogate model of blood haemodynamics for patient-specific carotid artery stenosis},
  author={Barzegar Gerdroodbary, Mostafa and Salavatidezfouli, Sajad},
  journal={Journal of the Royal Society Interface},
  volume={22},
  number={224},
  pages={20240774},
  year={2025},
  publisher={The Royal Society}
}

@article{du2022deep,
  title={{Deep learning-based surrogate model for three-dimensional patient-specific computational fluid dynamics}},
  author={Du, Pan and Zhu, Xiaozhi and Wang, Jian-Xun},
  journal={Physics of Fluids},
  volume={34},
  number={8},
  year={2022},
  publisher={AIP Publishing},
  doi={10.1063/5.0101128}
}

@book{aburahma2010noninvasive,
  title={Noninvasive peripheral arterial diagnosis},
  author={AbuRahma, Ali and Bergan, John},
  year={2010},
  publisher={Springer Science \& Business Media},
  ISBN={978-1-84882-954-1},
  doi={10.1007/978-1-84882-955-8}
}

@article{krankenberg2007nitinol,
  title={Nitinol stent implantation versus percutaneous transluminal angioplasty in superficial femoral artery lesions up to 10 cm in length: the femoral artery stenting trial (FAST)},
  author={Krankenberg, Hans and Schlüter, Michael and Steinkamp, Hermann J and Bürgelin, Karlheinz and Scheinert, Dierk and Schulte, Karl-Ludwig and Minar, Erich and Peeters, Patrick and Bosiers, Marc and Tepe, Gunnar and others},
  journal={Circulation},
  volume={116},
  number={3},
  pages={285--292},
  year={2007},
  publisher={Lippincott Williams \& Wilkins},
  doi={10.1161/CIRCULATIONAHA.107.689141}
}

@article{ohana2014detailed,
  title={Detailed cross-sectional study of 60 superficial femoral artery occlusions: morphological quantitative analysis can lead to a new classification},
  author={Ohana, Micka{\"e}l and El Ghannudi, Soraya and Girsowicz, Elie and Lejay, Anne and Georg, Yannick and Thaveau, Fabien and Chakfe, Nabil and Roy, Catherine},
  journal={Cardiovascular diagnosis and therapy},
  volume={4},
  number={2},
  pages={71},
  year={2014},
  doi={10.3978/j.issn.2223-3652.2014.01.01}
}

@article{choi2009methods,
  title={Methods for quantifying three-dimensional deformation of arteries due to pulsatile and nonpulsatile forces: implications for the design of stents and stent grafts},
  author={Choi, Gilwoo and Cheng, Christopher P and Wilson, Nathan M and Taylor, Charles A},
  journal={Annals of biomedical engineering},
  volume={37},
  number={1},
  pages={14--33},
  year={2009},
  publisher={Springer},
  doi={10.1016/j.jvir.2009.08.027}
}

@article{ninno2025silico,
  title={In Silico, Patient-Specific Assessment of Local Hemodynamic Predictors and Neointimal Hyperplasia Localisation in an Arteriovenous Graft},
  author={Ninno, Federica and Stokes, Catriona and Aboian, Edouard and Dardik, Alan and Strosberg, David and Balabani, Stavroula and D{\'\i}az-Zuccarini, Vanessa},
  journal={Annals of Biomedical Engineering},
  pages={1--15},
  year={2025},
  publisher={Springer},
  doi={10.1007/s10439-025-03737-8}
}

@article{Berkooz1993ThePOD,
  title={{The Proper Orthogonal Decomposition in the Analysis of Turbulent Flows}},
  author={Gal Berkooz and Philip Holmes and John L. Lumley},
  journal={Annual Review of Fluid Mechanics},
  year={1993},
  volume={25},
  pages={539-575},
  publisher={Annual Reviews 4139 El Camino Way, PO Box 10139, Palo Alto, CA 94303-0139, USA},
  doi={10.1146/annurev.fl.25.010193.002543}
}

@inproceedings{optuna_2019,
    title={Optuna: A Next-generation Hyperparameter Optimization Framework},
    author={Akiba, Takuya and Sano, Shotaro and Yanase, Toshihiko and Ohta, Takeru and Koyama, Masanori},
    booktitle={Proceedings of the 25th {ACM} {SIGKDD} International Conference on Knowledge Discovery and Data Mining},
    year={2019},
    doi={10.1145/3292500.3330701}
}

@article{hong2017characteristics,
  title={Characteristics of pulsatile flows in curved stenosed channels},
  author={Hong, Hyeonji and Yeom, Eunseop and Ji, Ho Seong and Kim, Hyun Dong and Kim, Kyung Chun},
  journal={PloS one},
  volume={12},
  number={10},
  pages={e0186300},
  year={2017},
  publisher={Public Library of Science San Francisco, CA USA},
  doi={10.1371/journal.pone.0186300}
}

@article{chatpattanasiri2023towards,
  title={{Towards Reduced Order Models via Robust Proper Orthogonal Decomposition to Capture Personalised Aortic Haemodynamics}},
  author={Chatpattanasiri, Chotirawee and Franzetti, Gaia and Bonfanti, Mirko and Diaz-Zuccarini, Vanessa and Balabani, Stavroula},
  journal={Journal of Biomechanics},
  volume={158},
  pages={111759},
  year={2023},
  publisher={Elsevier},
  doi={10.1016/j.jbiomech.2023.111759}
}

@article{altalhan2025imbalanced,
  title={Imbalanced data problem in machine learning: A review},
  author={Altalhan, Manahel and Algarni, Abdulmohsen and Alouane, Monia Turki-Hadj},
  journal={IEEE Access},
  year={2025},
  publisher={IEEE},
  doi={10.1109/ACCESS.2025.3531662}
}

@article{banerjee2023machine,
  title={Machine learning in rare disease},
  author={Banerjee, Jineta and Taroni, Jaclyn N and Allaway, Robert J and Prasad, Deepashree Venkatesh and Guinney, Justin and Greene, Casey},
  journal={Nature Methods},
  volume={20},
  number={6},
  pages={803--814},
  year={2023},
  publisher={Nature Publishing Group US New York},
  doi={10.1038/s41592-023-01886-z}
}

@article{yao2024image2flow,
  title={Image2Flow: A proof-of-concept hybrid image and graph convolutional neural network for rapid patient-specific pulmonary artery segmentation and CFD flow field calculation from 3D cardiac MRI data},
  author={Yao, Tina and Pajaziti, Endrit and Quail, Michael and Schievano, Silvia and Steeden, Jennifer and Muthurangu, Vivek},
  journal={PLOS Computational Biology},
  volume={20},
  number={6},
  pages={e1012231},
  year={2024},
  publisher={Public Library of Science San Francisco, CA USA},
  doi={10.1371/journal.pcbi.1012231}
}

@article{bhatti2023deep,
  title={Deep learning with graph convolutional networks: An overview and latest applications in computational intelligence},
  author={Bhatti, Uzair Aslam and Tang, Hao and Wu, Guilu and Marjan, Shah and Hussain, Aamir},
  journal={International Journal of Intelligent Systems},
  volume={2023},
  number={1},
  pages={8342104},
  year={2023},
  publisher={Wiley Online Library},
  doi={10.1155/2023/8342104}
}

@article{pegolotti2024learning,
  title={Learning reduced-order models for cardiovascular simulations with graph neural networks},
  author={Pegolotti, Luca and Pfaller, Martin R and Rubio, Natalia L and Ding, Ke and Brufau, Rita Brugarolas and Darve, Eric and Marsden, Alison L},
  journal={Computers in Biology and Medicine},
  volume={168},
  pages={107676},
  year={2024},
  publisher={Elsevier},
  doi={10.1016/j.compbiomed.2023.107676}
}

@article{chauhan2024brief,
  title={A brief review of hypernetworks in deep learning},
  author={Chauhan, Vinod Kumar and Zhou, Jiandong and Lu, Ping and Molaei, Soheila and Clifton, David A},
  journal={Artificial Intelligence Review},
  volume={57},
  number={9},
  pages={250},
  year={2024},
  publisher={Springer},
  doi={10.1007/s10462-024-10862-8}
}

@article{pan2023NIF,
  title={Neural implicit flow: a mesh-agnostic dimensionality reduction paradigm of spatio-temporal data},
  author={Pan, Shaowu and Brunton, Steven L and Kutz, J Nathan},
  journal={Journal of Machine Learning Research},
  volume={24},
  number={41},
  pages={1--60},
  year={2023},
  url={http://jmlr.org/papers/v24/22-0365.html}
}

@article{cai2021physics,
  title={Physics-informed neural networks (PINNs) for fluid mechanics: A review},
  author={Cai, Shengze and Mao, Zhiping and Wang, Zhicheng and Yin, Minglang and Karniadakis, George Em},
  journal={Acta Mechanica Sinica},
  volume={37},
  number={12},
  pages={1727--1738},
  year={2021},
  publisher={Springer},
  doi={10.1007/s10409-021-01148-1}
}

@article{csala2025physics,
  title={Physics-constrained coupled neural differential equations for one dimensional blood flow modeling},
  author={Csala, Hunor and Mohan, Arvind and Livescu, Daniel and Arzani, Amirhossein},
  journal={Computers in Biology and Medicine},
  volume={186},
  pages={109644},
  year={2025},
  publisher={Elsevier},
  doi={10.1016/j.compbiomed.2024.109644}
}

@article{wong2024strategies,
  title={Strategies for multi-case physics-informed neural networks for tube flows: a study using 2D flow scenarios},
  author={Wong, Hong Shen and Chan, Wei Xuan and Li, Bing Huan and Yap, Choon Hwai},
  journal={Scientific Reports},
  volume={14},
  number={1},
  pages={11577},
  year={2024},
  publisher={Nature Publishing Group UK London},
  doi={/10.1038/s41598-024-62117-9}
}

@article{chen2021physics,
  title={Physics-informed machine learning for reduced-order modeling of nonlinear problems},
  author={Chen, Wenqian and Wang, Qian and Hesthaven, Jan S and Zhang, Chuhua},
  journal={Journal of computational physics},
  volume={446},
  pages={110666},
  year={2021},
  publisher={Elsevier},
  doi={10.1016/j.jcp.2021.110666}
}

@article{fu2023physics,
  title={Physics-data combined machine learning for parametric reduced-order modelling of nonlinear dynamical systems in small-data regimes},
  author={Fu, Jinlong and Xiao, Dunhui and Fu, Rui and Li, Chenfeng and Zhu, Chuanhua and Arcucci, Rossella and Navon, Ionel M},
  journal={Computer Methods in Applied Mechanics and Engineering},
  volume={404},
  pages={115771},
  year={2023},
  publisher={Elsevier},
  doi={10.1016/j.cma.2022.115771}
}

@article{chan2025role,
  title={Role of physics-informed constraints in real-time estimation of 3D vascular fluid dynamics using multi-case neural network},
  author={Chan, Wei Xuan and Ding, Wenhao and Li, Binghuan and Wong, Hong Shen and Yap, Choon Hwai},
  journal={Computers in Biology and Medicine},
  volume={190},
  pages={110074},
  year={2025},
  publisher={Elsevier},
  doi={10.1016/j.compbiomed.2025.110074}
}

@article{huo2018acute,
  title={Acute tachycardia increases aortic distensibility, but reduces total arterial compliance up to a moderate heart rate},
  author={Huo, Yunlong and Chen, Huan and Kassab, Ghassan S},
  journal={Frontiers in Physiology},
  volume={9},
  pages={1634},
  year={2018},
  publisher={Frontiers Media SA},
  doi={10.3389/fphys.2018.01634}
}

@article{williamson2024literature,
  title={Literature survey for in-vivo reynolds and womersley numbers of various arteries and implications for compliant in-vitro modelling},
  author={Williamson, Petra N and Docherty, Paul D and Jermy, Mark and Steven, Briana M},
  journal={Cardiovascular Engineering and Technology},
  volume={15},
  number={4},
  pages={418--430},
  year={2024},
  publisher={Springer},
  doi={10.1007/s13239-024-00723-4}
}

@article{nader2019blood,
  title={Blood rheology: key parameters, impact on blood flow, role in sickle cell disease and effects of exercise},
  author={Nader, Elie and Skinner, Sarah and Romana, Marc and Fort, Romain and Lemonne, Nathalie and Guillot, Nicolas and Gauthier, Alexandra and Antoine-Jonville, Sophie and Renoux, C{\'e}line and Hardy-Dessources, Marie-Dominique and others},
  journal={Frontiers in physiology},
  volume={10},
  pages={1329},
  year={2019},
  publisher={Frontiers Media SA},
  doi={10.3389/fphys.2019.01329}
}

@misc{outflowBC,
  title = {{Ansys Fluent} user guide: Outflow Boundary Conditions},
  howpublished = {\url{https://www.afs.enea.it/project/neptunius/docs/fluent/html/ug/node247.htm}},
  note = {Accessed: 2025-10-30}
}

@article{buoso2019reduced,
  title={{Reduced-order modeling of blood flow for noninvasive functional evaluation of coronary artery disease}},
  author={Buoso, Stefano and Manzoni, Andrea and Alkadhi, Hatem and Plass, Andr{\'e} and Quarteroni, Alfio and Kurtcuoglu, Vartan},
  journal={Biomechanics and modeling in mechanobiology},
  volume={18},
  pages={1867--1881},
  year={2019},
  publisher={Springer},
  doi={10.1007/s10237-019-01182-w}
}

\end{document}


\maketitle

\section{Introduction}

This document provides a detailed description of the methodology followed to generate the dataset underpinning the reduced-order modelling and machine learning (ML) analyses reported in the main article. 

This supplementary material expands on four key aspects. Section \ref{sec:Baseline geometry} describes the construction and validation of the baseline cylindrical model that served as the template for all subsequent deformations. Section \ref{sec:Synthetic geometry generation} outlines the parameterisation of vascular variations and the mathematical derivations of the transformations used to introduce stenosis, tapering, and curvature. Section \ref{sec:Synthetic flowrate waveform generation} details the application of principal component analysis (PCA) to patient-derived inflow waveforms and the procedure for synthesising physiologically realistic inflow conditions. Section \ref{sec:CFD} presents the computational fluid dynamics (CFD) setup, governing equations, and boundary conditions. Validation of these CFD results are given in Section \ref{sec:CFD Validation}. Finally, Section \ref{sec:Critical regions} explains how haemodynamic indices were computed and thresholded to define critical, prone to restenosis, regions. Together, these descriptions ensure transparency, reproducibility, and completeness of the dataset employed in the main study.

\begin{figure}[ht]
\centering
\includegraphics[width=\textwidth]{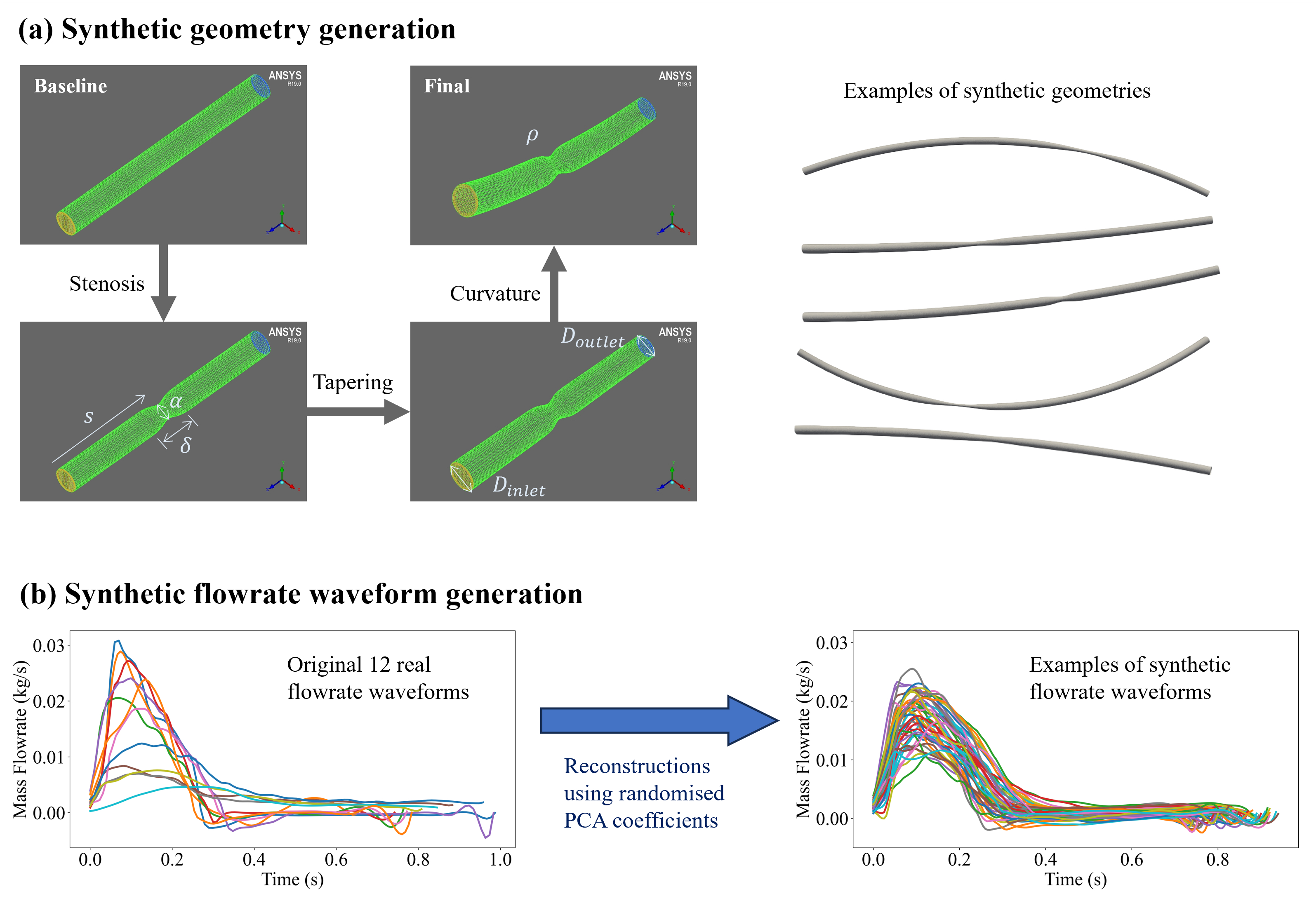}
\caption{
Dataset generation. (a) Synthetic geometries are created by deforming a baseline model in three steps, using six geometric parameters. Examples of generated geometries are shown. (b) Synthetic flowrate waveforms obtained by applying PCA to 12 real flowrate waveforms and randomising coefficients. Examples of original and synthetic waveforms are shown.}
\label{fig:dataset_generation}
\end{figure}

\begin{table}[ht]
\centering
\caption{Definition and sampling ranges of geometric parameters used for synthetic geometry generation.}
\label{tab:geo_parameters}
\begin{tabular}{llll}
\hline
\textbf{Parameter} & \textbf{Symbol} & \textbf{Definition} & \textbf{Range} \\
\hline
Inlet diameter      & $D_{\text{inlet}}$ & Vessel diameter at inlet cross-section          & 4.8 -- 7.2 mm \\
Outlet diameter     & $D_{\text{outlet}}$ & Vessel diameter at outlet cross-section        & 4.3 -- 6.5 mm \\
Stenosis severity   & $\alpha$     & Maximum ratio of reduction in diameter         & 0.2 -- 0.7 \\
Stenosis length     & $\delta$     & The length of stenosis                         & 14.4 -- 66.4 mm \\
Stenosis position   & $s$          & Axial location of stenosis centre              & 120 -- 210 mm \\
Curvature           & $\rho$       & Vessel curvature (inverse radius)              & -0.004 -- 0.004 mm$^{-1}$ \\
\hline
\end{tabular}
\end{table}

\section{Baseline geometry} \label{sec:Baseline geometry}
The dataset generation process began with the construction of a baseline geometry which is a simple cylinder with a diameter $D_0=6$ m and length $L=300$ mm. A Cartesian coordinate system is set with its origin at the inlet cross-section centre, and the $y$-axis aligned to the cylinder’s length. Volumetric meshes were created for this geometry using ICEM CFD 19.0. Ansys Fluent 19.0 (Ansys Inc., PA, USA) was employed to perform CFD simulations for a mesh-independence analysis. 

\begin{figure}[ht]
\centering
\includegraphics[width=0.8\textwidth]{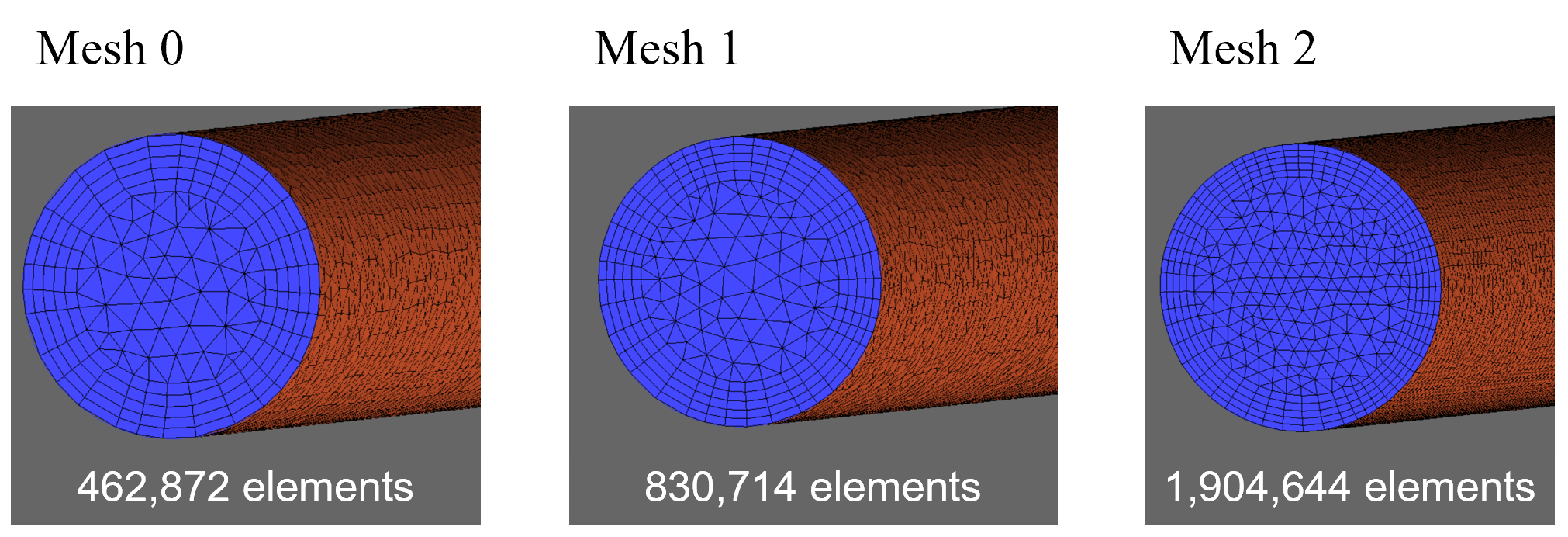}
\caption{Meshes used for the mesh-independence study: Mesh 0 (coarse), Mesh 1 (medium), and Mesh 2 (fine). The number of elements in each mesh is shown.}
\label{fig:Baseline_mesh_indepence}
\end{figure}

The mesh-independence analysis was carried out by comparing three progressively refined meshes against the analytical Poiseuille solution under steady-state Newtonian flow, with blood viscosity $\mu = 0.0032 \,\text{kg}/(\text{m}\cdot\text{s})$ and density $\rho = 1056 \,\text{kg}/\text{m}^3$. A uniform inlet velocity of $0.01 \,\text{m/s}$ was prescribed, with a pressure outlet condition set to zero. Three meshes of increasing resolution were tested, as shown in Figure \ref{fig:Baseline_mesh_indepence}.

\begin{figure}[ht]
\centering
\includegraphics[width=0.7\textwidth]{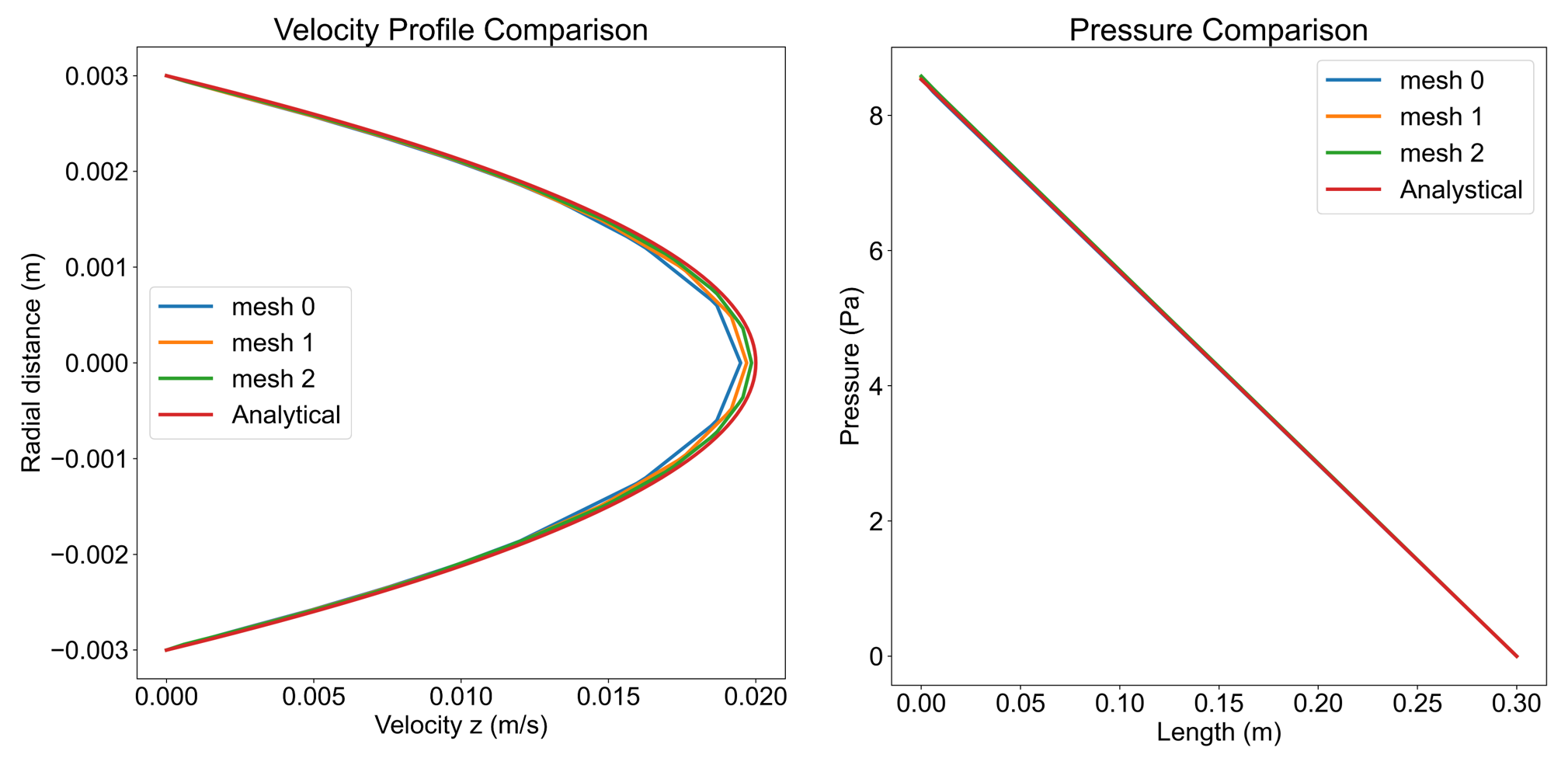}
\caption{Mesh-independence result.}
\label{fig:Mesh_independence_result}
\end{figure}


The velocity profiles (Figure \ref{fig:Mesh_independence_result}, left) and pressure distributions (Figure \ref{fig:Mesh_independence_result}, right) showed excellent agreement across all meshes, with deviations decreasing as resolution increased. Mesh 0 slightly underestimated the centreline velocity (-2.55\% error) and pressure drop, while Mesh 2 provided the closest agreement to the analytical solution (-0.75\% velocity error). However, the improvement from Mesh 1 to Mesh 2 was marginal compared to the significant increase in computational cost (over twice as many elements). Mesh 1 achieved a good compromise, with velocity 1.55\% and pressure drop deviation below 0.3\%. Therefore, Mesh 1 was selected as the baseline mesh for all subsequent deformations.

\section{Synthetic geometry generation} \label{sec:Synthetic geometry generation}
The baseline mesh was systematically deformed to create a diverse population of stenosed femoral artery geometries. Six parameters (Table \ref{tab:geo_parameters}) were defined to control anatomical variation: inlet diameter ($D_{\text{inlet}}$), outlet diameter ($D_{\text{outlet}}$), stenosis severity ($\alpha$), stenosis length ($\delta$), stenosis position ($s$), and curvature ($\rho$). These parameters respectively governed tapering, overall narrowing, location, severity and length of the lesion, and vessel bending. Three sequential transformations were implemented to modify mesh node coordinates.

\begin{enumerate}
    \item \textbf{Stenosis}:  
    A focal narrowing was introduced at axial location $s$ with severity $\alpha$ and spread $\delta$.  
    The cross-sectional scaling followed a Gaussian bell-shaped curve\footnote{Similar approach was used to create stenosis in vessels such as in Buoso et al. \cite{buoso2019reduced} and Wong et al. \cite{wong2024strategies}.}, rearranged to be expressed directly in terms of $\alpha$ and $\delta$:
    \begin{subequations}\label{eq:stenosis}
    \begin{align}
        x' &= x \cdot \frac{D_{\text{inlet}}}{D_0}\left(1 - \alpha \exp\left[\left(\frac{z-s}{\delta}\right)^2 \cdot 4 \ln\left(\frac{0.10}{\alpha}\right)\right]\right), \label{eq:stenosis_x} \\
        y' &= y \cdot \frac{D_{\text{inlet}}}{D_0}\left(1 - \alpha \exp\left[\left(\frac{z-s}{\delta}\right)^2 \cdot 4 \ln\left(\frac{0.10}{\alpha}\right)\right]\right), \label{eq:stenosis_y} \\
        z' &= z. \label{eq:stenosis_z}
    \end{align}
    \end{subequations}
    Here, the constant $0.10$ was selected so that the stenosis length $\delta$ is defined from the points where the diameter is equal to $90\%$ of $D_{\text{inlet}}$. Detail derivations for these equations are presented in Appendix \ref{appendix:A}.

    \item \textbf{Tapering}:  
    A gradual change in diameter from $D_{\text{inlet}}$ to $D_{\text{outlet}}$ was imposed along the vessel length $L_0$, applied through a linear scaling factor in $z$:
    \begin{subequations}\label{eq:tapering}
    \begin{align}
        x'' &= x' \left(\frac{D_{\text{outlet}}/D_{\text{inlet}} - 1}{L_0} z' + 1\right), \label{eq:tapering_x} \\
        y'' &= y' \left(\frac{D_{\text{outlet}}/D_{\text{inlet}} - 1}{L_0} z' + 1\right), \label{eq:tapering_y} \\
        z'' &= z'. \label{eq:tapering_z}
    \end{align}
    \end{subequations}

    \item \textbf{Curvature}:  
    A uniform curvature was introduced using the parameter $\rho$. The radius of curvature was defined as $R = \tfrac{1}{\rho} - y''$, and the angular displacement $\theta = z'' \rho$. The transformed coordinates were given by
    \begin{subequations}\label{eq:curvature}
    \begin{align}
        x''' &= x'', \label{eq:curvature_x} \\
        y''' &= y'' + R \big(1 - \cos(\theta)\big), \label{eq:curvature_y} \\
        z''' &= R \sin(\theta). \label{eq:curvature_z}
    \end{align}
    \end{subequations}
    This yielded geometries with different degrees of bending. The derivations for these equations can be found in Appendix \ref{appendix:B}.
\end{enumerate}

To efficiently sample physiologically plausible configurations, parameter values were drawn using Latin Hypercube Sampling (LHS) within clinically realistic ranges \cite{aburahma2010noninvasive, krankenberg2007nitinol, choi2009methods, ohana2014detailed}. For each generated geometry, the mesh file was rewritten with updated nodal coordinates, and parameter values were stored in a metadata file for traceability. In total, 500 distinct geometries were produced. Examples are shown in Figure \ref{fig:dataset_generation}a.

\section{Synthetic flowrate waveform generation} \label{sec:Synthetic flowrate waveform generation}
A set of 12 patient-derived femoral artery inflow waveforms acquired by Doppler-ultrasound and reported in Ninno et al. \cite{Ninno2024modelling_lower_limb} and Chatpattnasiri et al. \cite{chatpattanasiri2025ML} were used to form the base population. Each waveform, represented as a discrete time series of mass flowrate, was arranged into a matrix  

\begin{equation}
X^{(w)} = \begin{bmatrix}
| & | & & | \\
\mathbf{x}^{(w)}_1 & \mathbf{x}^{(w)}_2 & \cdots & \mathbf{x}^{(w)}_{12} \\
| & | & & |
\end{bmatrix},
\label{eq:snapshot_matrix_w}
\end{equation}

where $\mathbf{x}^{(w)}_i \in \mathbb{R}^n$ is the waveform for patient $i$, sampled over one cardiac cycle, and $n=101$ since all real waveforms were uniformly resampled at 101 points regardless of their original period.

PCA was then applied via SVD:  

\begin{equation}
X^{(w)} = U^{(w)} \Sigma^{(w)} {V^{(w)}}^\top,
\label{eq:svd_w}
\end{equation}

where $U^{(w)}$ contains temporal modes, $\Sigma^{(w)}$ is the diagonal matrix of singular values, and $V^{(w)}$ contains the PCA coefficients. Each waveform can be reconstructed as  

\begin{equation}
\mathbf{x}^{(w)}_i \approx U^{(w)} \, \mathbf{c}^{(w)}_i,
\label{eq:reconstruction_w}
\end{equation}

with coefficients  

\begin{equation}
\mathbf{c}^{(w)}_i = \Sigma^{(w)} {V^{(w)}}^\top(:,i).
\label{eq:coeff_w}
\end{equation}

To generate synthetic waveforms, we sampled new coefficient sets $\mathbf{c}^{(w)}_{\text{syn}}$ using LHS within the range  

\begin{equation}
c^{(w),\text{syn}}_{j} \in \big[ \mu^{(w)}_j - \sigma^{(w)}_j, \, \mu^{(w)}_j + \sigma^{(w)}_j \big],
\label{eq:lhs_w}
\end{equation}

where $\mu^{(w)}_j$ and $\sigma^{(w)}_j$ are the mean and standard deviation of coefficient $j$ across the base population. This ensures that synthetic waveforms remain within $\pm 1$ SD of the observed variability while covering the multidimensional parameter space more uniformly than random sampling.  

Each synthetic inflow was reconstructed as  

\begin{equation}
\mathbf{x}^{(w)}_{\text{syn}} = U^{(w)} \, \mathbf{c}^{(w)}_{\text{syn}}.
\label{eq:syn_waveform_w}
\end{equation}

The waveform period $T^{(w)}$ was varied in the range $\mu^{(w)}_T$ $\pm \sigma^{(w)}_T$:  

\begin{equation}
T^{(w)}_{\text{syn}} \in \big[ \mu^{(w)}_T - \sigma^{(w)}_T, \, \mu^{(w)}_T + \sigma^{(w)}_T \big].
\label{eq:period_w}
\end{equation}

The resulting waveforms were resampled onto a uniform time grid with step size $\Delta t = 0.005 \,\text{s}$.  

This procedure yielded 500 unique synthetic flowrate waveforms that retained realistic systolic peaks, diastolic decays, and pulsatile periodicity. Examples of both base and generated profiles are shown in Figure \ref{fig:dataset_generation}b.  

To confirm physiological relevance, Reynolds and Womersley numbers were computed using the inlet diameter $D_{\text{inlet}}$, and assuming a value of blood viscosity of $\mu = 0.004~\mathrm{Pa\cdot s}$ for all cases \cite{nader2019blood}. The Reynolds number based on time-averaged flow rate ranged from $95.5$ to $270.0$, while the peak Reynolds number (using the cycle-maximum flow rate)  between $287.7$ and $1026.0$. The Womersley number, also using $D_{\text{inlet}}$, varies from $6.52$ to $10.72$. These ranges are consistent with values reported in the literature for femoral arteries \cite{huo2018acute, williamson2024literature}.

\section{CFD}  \label{sec:CFD}



Blood flow in each geometry–waveform pair was simulated using CFD to generate the haemodynamic indices dataset. The CFD simulations were run using Ansys Fluent 19.0 (Ansys Inc., PA, USA) on the UCL Department of Computer Science cluster (Intel Xeon Gold 5118 at 2.3 GHz, 10 processors). The CFD methodology followed that of Ninno et al. \cite{Ninno2024modelling_lower_limb}. Blood was modelled as a non-Newtonian fluid with Carreau viscosity and constant density. The governing equations consisted of the incompressible Navier–Stokes and continuity equations:

\begin{subequations}\label{eq:NS_cont}
\begin{align}
\rho \left( \frac{\partial \mathbf{u}}{\partial t} + \mathbf{u} \cdot \nabla \mathbf{u} \right) 
&= -\nabla p + \nabla \cdot \left[ \mu(\dot{\gamma}) \left( \nabla \mathbf{u} + (\nabla \mathbf{u})^\top \right) \right], 
\label{eq:NS_momentum} \\
\nabla \cdot \mathbf{u} &= 0,
\label{eq:NS_continuity}
\end{align}
\end{subequations}

where $\mathbf{u}$ is the velocity vector, $p$ is the pressure, and $\rho = 1056 \,\text{kg}/\text{m}^3$. The Carreau viscosity is given by:

\begin{equation}
\mu(\dot{\gamma}) = \mu_\infty + \left( \mu_0 - \mu_\infty \right) \left[ 1 + \left( \lambda \dot{\gamma} \right)^2 \right]^{\tfrac{n-1}{2}},
\label{eq:carreau}
\end{equation}

with zero shear viscosities $\mu_0 = 0.25 \,\text{kg/(m·s)}$, infinite shear viscosities $\mu_\infty = 0.0035 \,\text{kg/(m·s)}$, time constant $\lambda = 25 \,\text{s}$, and the power-law index $n = 0.25$.

At the inlet, a time-dependent parabolic velocity profile was prescribed by scaling the synthetic flowrate waveform to the inlet cross-sectional area. The outlet boundary was treated as outflow \cite{outflowBC}, and the vessel wall was assumed rigid with no-slip boundary conditions.

Numerical discretisation employed a finite volume method with a second-order upwind scheme for momentum. Time integration was performed using a second-order implicit scheme with a fixed time step of $\Delta t = 0.005 ,\text{s}$, consistent with the synthetic waveform resampling. Simulations were advanced for three cardiac cycles to eliminate transient effects, and results from the final cycle were extracted for analysis.

To generate the dataset of 500 cases, simulations were divided into 10 batches, each requiring 20–60 hours depending on the case setup and cluster resources. After completion, results were organised, compressed, and transferred to a local workstation, which added another 20–25 hours. Several stages overlapped. For instance, new CFD runs were initiated as soon as the previous batch had been prepared for download. Occasional cluster failures required reruns and prolonged the process. Overall, dataset generation consumed about 25 days (600 hours). To support the subsequent ROM and ML development phases, the dataset was initially planned for a 300:100:100 training-validation-test split. However, four cases continued to fail at the final rerun stage and were excluded, resulting in 469 usable cases and adjusted splits of 300:97:99 for training, validation and testing, respectively.

\section{CFD Validation}  \label{sec:CFD Validation}

\subsection{Comparisons against published results}  \label{ssec:Comparison with a past study}

\begin{figure}
    \centering
    \includegraphics[width=0.8\textwidth]{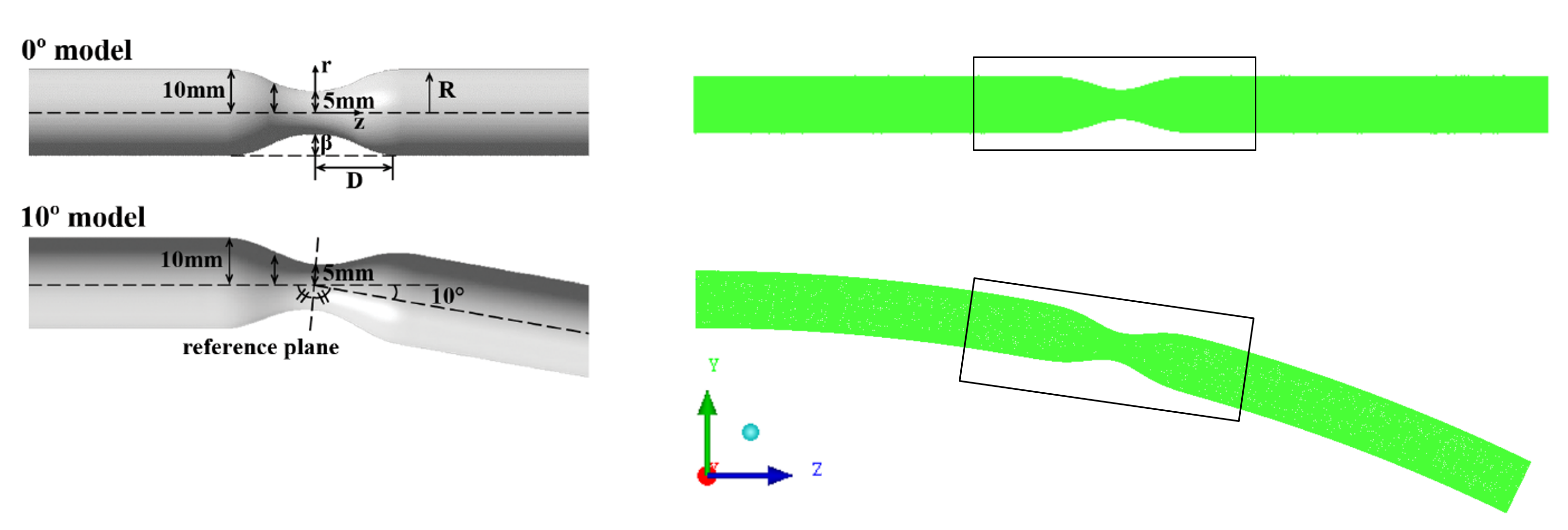}
    \caption{Geometrical setup for validation cases based on Hong et al. \cite{hong2017characteristics}. 
    Left: reference geometries reported by Hong et al. \cite{hong2017characteristics}. 
    Right: corresponding geometries generated in the present study. Two cases were considered: a straight stenosed channel ($0^\circ$ model) and a curved stenosed channel ($10^\circ$ model).}
    \label{fig:validation1_geometry}
\end{figure}

\begin{figure}
    \centering
    \includegraphics[width=0.8\textwidth]{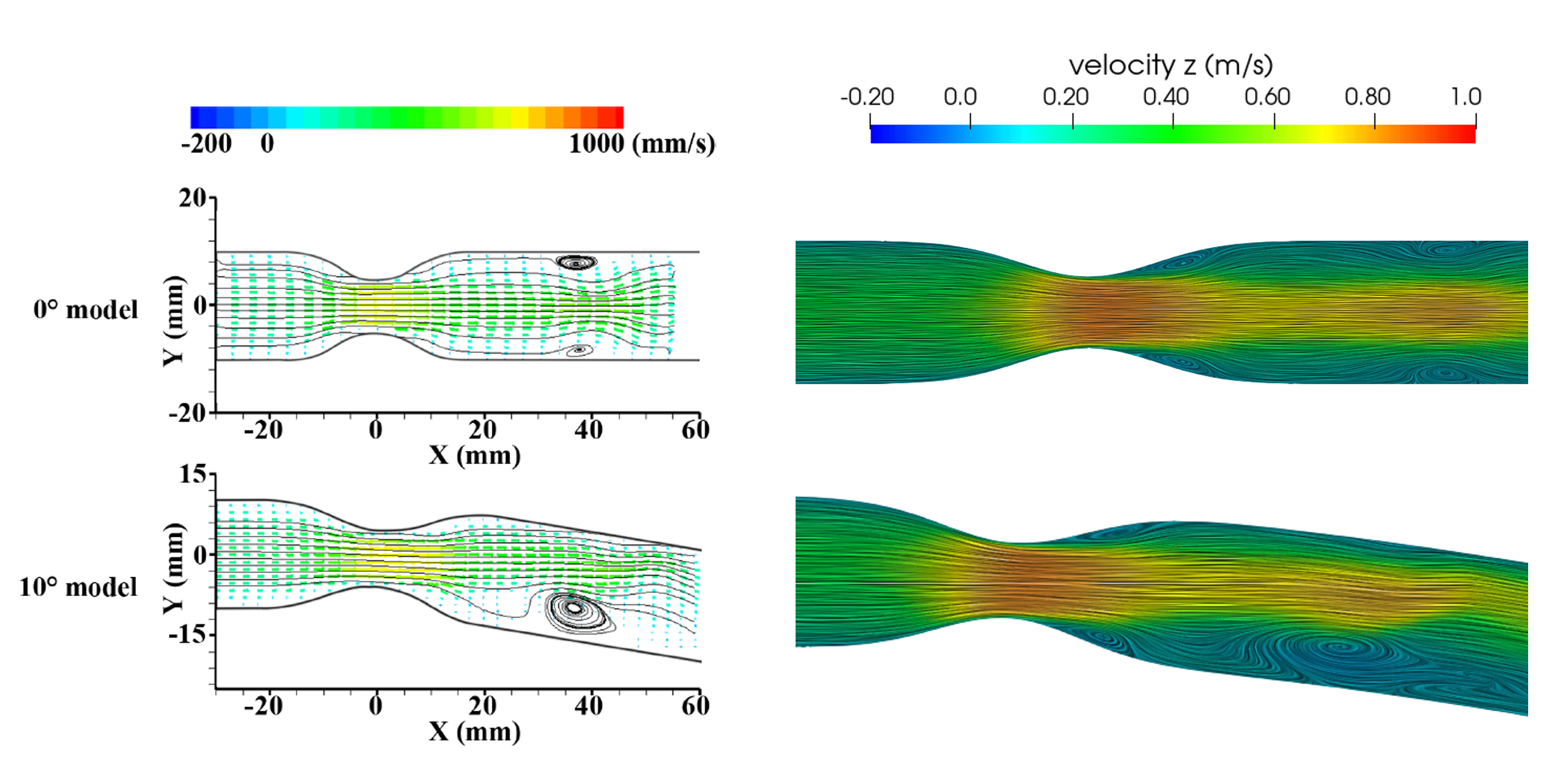}
    \caption{Comparison of velocity field and streamlines between Hong et al. \cite{hong2017characteristics} (left) and the present CFD simulations (right). 
    Results are shown for the straight stenosed channel ($0^\circ$ model) and the curved stenosed channel ($10^\circ$ model).}
    \label{fig:validation1_peak_velocity}
\end{figure}

To validate the CFD setup, additional simulations were performed and compared with the study of Hong et al. \cite{hong2017characteristics}, who investigated pulsatile flow in straight and curved stenosed channels. Two validation geometries were created: (i) a straight pipe with a central stenosis and (ii) a curved pipe with the same stenosis. While an effort was made to match the reported dimensions, exact replication was not possible because Hong et al. \cite{hong2017characteristics} employed different mathematical formulations to create the stenosis and bending. Their stenosis was described by a cosine function rather than our Gaussian-type curve, and their curvature was imposed as a sharp bend instead of a gradual arch. For both models, the inlet flowrate waveform was prescribed to reproduce the pulsatile profile reported in \cite{hong2017characteristics}. 

Figure \ref{fig:validation1_peak_velocity} compares the velocity fields obtained from our simulations with those of Hong et al. \cite{hong2017characteristics}. The flow structures, including jet acceleration through the stenosis and the recirculation zones in the post-stenotic region, exhibit good qualitative agreement. Minor discrepancies are visible in the size and strength of vortices. This can be attributed to differences in the exact geometries and fluid viscosity, but overall the results confirm that the present CFD setup reproduces the main haemodynamic features observed in previous validated studies.

\subsection{Internal consistency checks} \label{ssec:Internal consistency checks}  

\begin{figure}
    \centering
    \includegraphics[width=\textwidth]{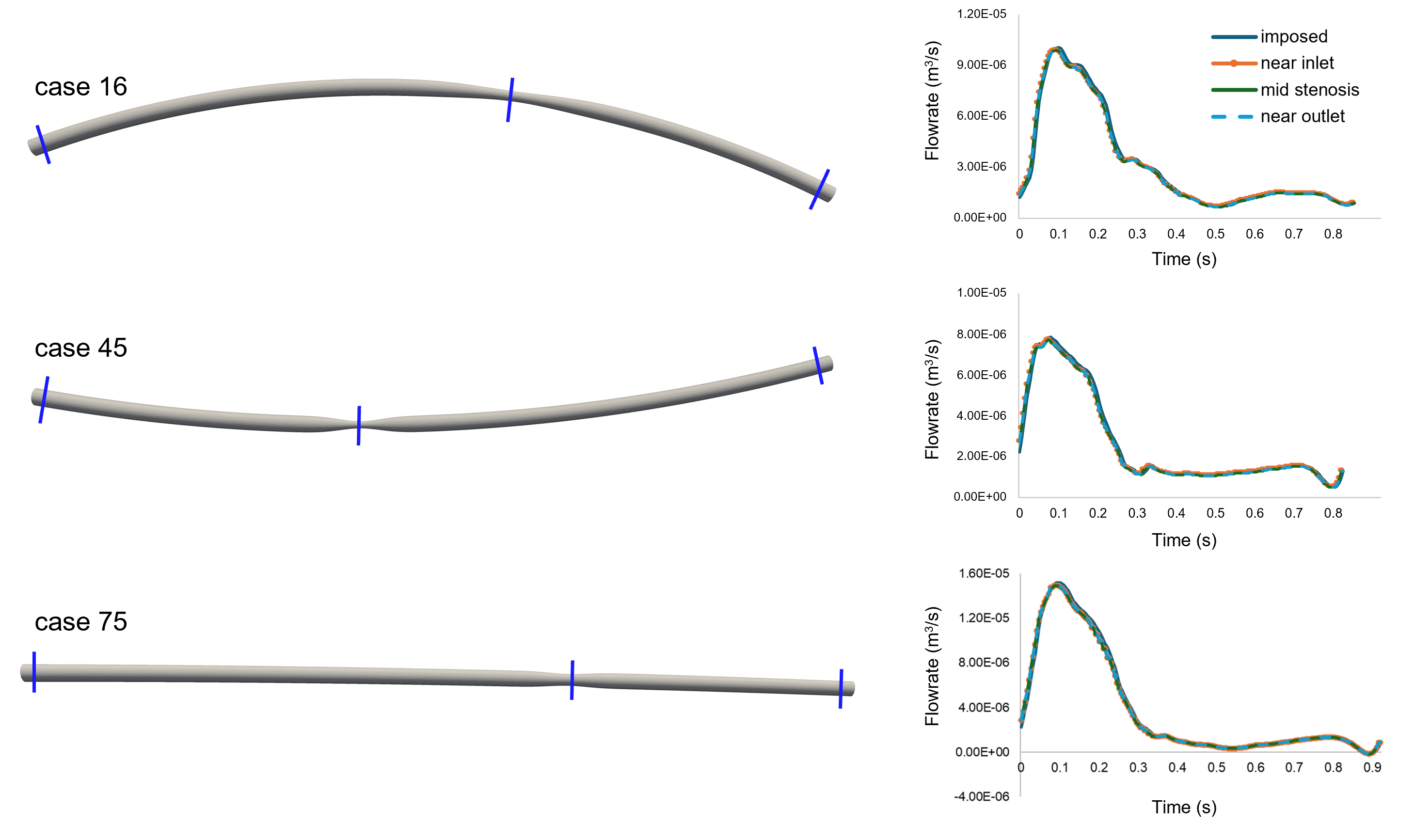}
    \caption{Mass conservation check for three randomly selected cases. Left: locations of measurement planes (blue lines) placed near the inlet, at the mid-stenosis, and near the outlet. Right: comparison of flowrate waveforms at the three planes (coloured curves) with the imposed inflow waveform.}
    \label{fig:validation2_flowrate}
\end{figure}

\begin{figure}
    \centering
    \includegraphics[width=\textwidth]{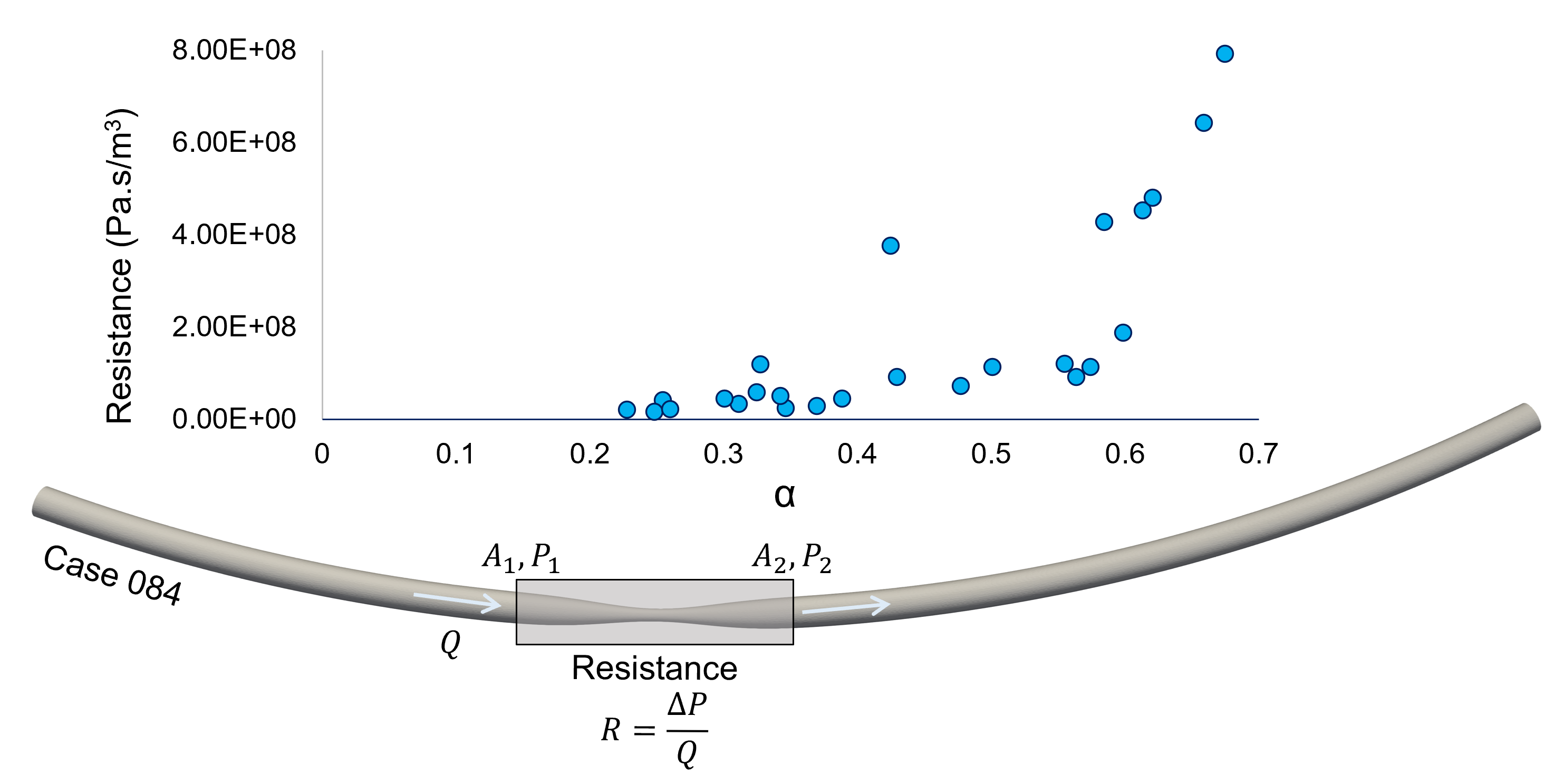}
    \caption{Flow resistance versus stenosis severity. Top: resistance $R = \Delta P / Q$ plotted against stenosis severity $\alpha$ for 25 random cases. Bottom: schematic illustrating the definition of resistance based on pressure drop across the stenosis and the time-averaged flowrate.}
    \label{fig:validation2_resistance}
\end{figure}

\begin{figure}
    \centering
    \includegraphics[width=\textwidth]{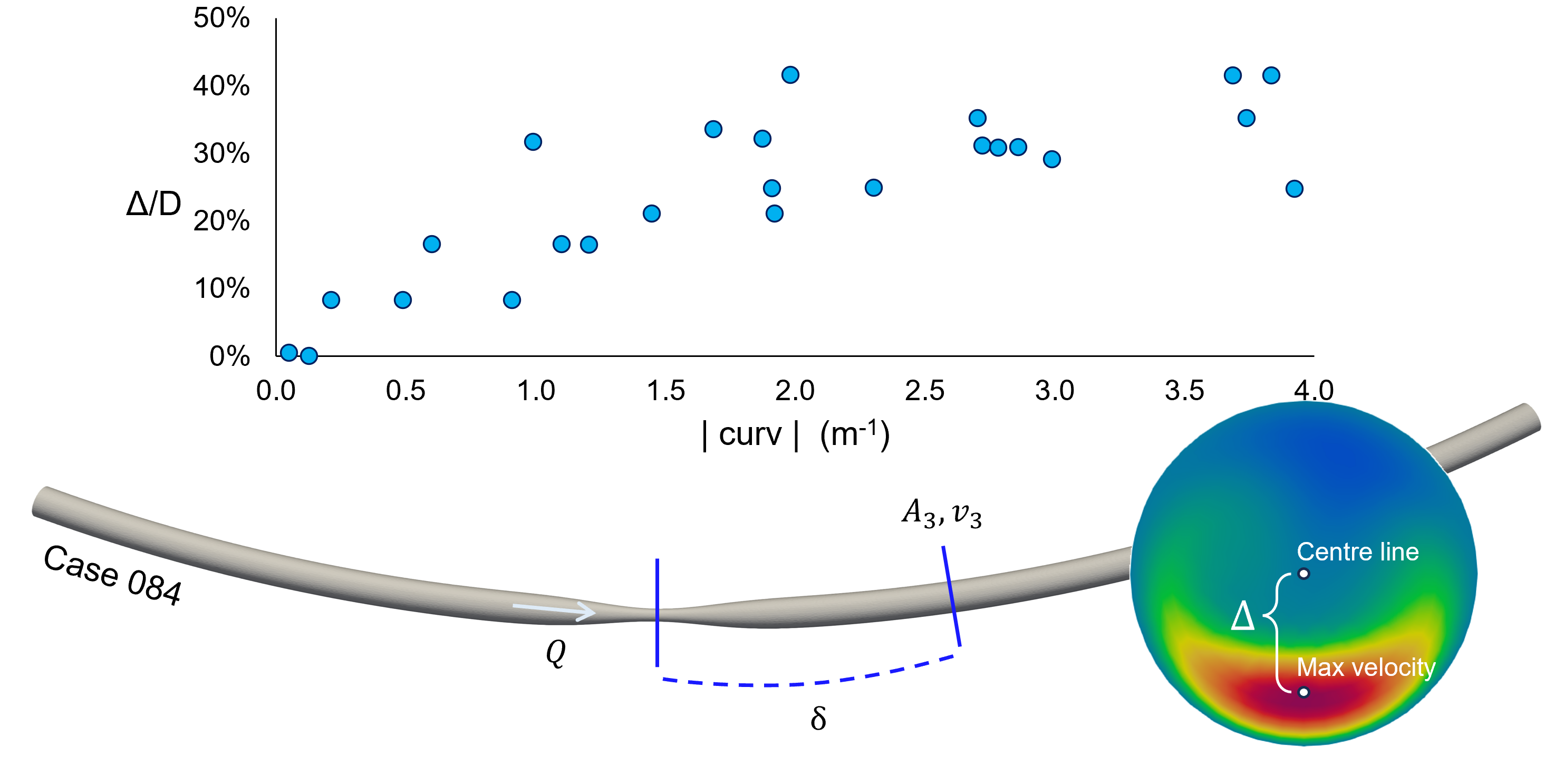}
    \caption{Velocity axis deviation versus curvature. Top: normalised deviation $\Delta/D$ of the maximum velocity location from the vessel centreline plotted against absolute curvature $|\rho|$ for 25 random cases. Bottom: schematic illustrating the measurement of $\Delta$ at a cross-section one stenosis length downstream of the throat. Bottom-right: velocity magnitude at the cross-section showing centreline and maximum velocity location.}
    \label{fig:validation2_deviation}
\end{figure}

To further assess the reliability of the generated dataset, additional validation tests were performed on randomly selected cases:  

\begin{enumerate}
    \item \textbf{Mass flowrate conservation.}  
    The flowrate waveforms were evaluated at three cross-sections: near the inlet, at the mid-stenosis, and near the outlet. As shown in Figure \ref{fig:validation2_flowrate}, the waveforms obtained from CFD closely reproduced the imposed inflow profiles across three randomly selected cases, with an average discrepancy below $5\%$. This confirms that mass conservation was satisfied and the boundary conditions were properly implemented.

    \item \textbf{Flow resistance versus stenosis severity.}  
    The flow resistance was evaluated across the stenosis as the ratio of mean pressure drop to the time-averaged flowrate. Figure \ref{fig:validation2_resistance} shows the results for 25 random cases, plotted against stenosis severity $\alpha$. Resistance increases with $\alpha$, although the relationship is nonlinear and the points are scattered due to other influencing parameters such as stenosis length, tapering, and curvature. Nevertheless, the overall trend confirms that the CFD results show the expected relationship between flow resistance and $\alpha$.  

    \item \textbf{Velocity axis deviation versus curvature.}  
    At a cross-section located one stenosis length downstream of the throat, the offset $\Delta$ between the vessel centreline and the location of maximum axial velocity was computed. Figure \ref{fig:validation2_deviation} plots this deviation (normalised by diameter) against absolute curvature $|\rho|$ for 25 random cases. The deviation shows variability across cases but generally increases with curvature magnitude, and is nearly absent when curvature is close to zero. This behaviour arises from the combined influence of curvature-induced secondary Dean vortices and the post-stenotic jet (similar to structures in Figure \ref{fig:validation1_peak_velocity}), which together redistribute the velocity profile asymmetrically. Overall, this trend is consistent with the expected flow behaviour in curved stenosed vessels.
\end{enumerate}  

Together, these checks confirm that the dataset exhibits consistent haemodynamic behaviour aligned with physical expectations. Mass conservation was strictly satisfied, flow resistance increased with stenosis severity, and velocity axis deviation correlated with curvature. These trends provide additional confidence in the fidelity of the generated cases.

\section{Identification of critical, restenosis prone,  regions}  \label{sec:Critical regions}

For all 496 usable cases, the WSS was computed directly from the velocity gradients at the wall. From the instantaneous WSS field, TAWSS and OSI were obtained using:

\begin{subequations}
\begin{align}
\text{TAWSS} = \frac{1}{T} \int_{0}^{T} |\boldsymbol{\tau}_w| \, dt
\label{eq:TAWSS} \\
\text{OSI} = 0.5 \left(1 - \frac{|\int_{0}^{T} \boldsymbol{\tau}_w \, dt|}{\int_{0}^{T} |\boldsymbol{\tau}_w| \, dt}\right)
\label{eq:OSI}
\end{align}
\end{subequations}

where $\boldsymbol{\tau}_w$ is the instantaneous wall shear stress vector and $T$ is the cardiac cycle period.

Critical regions were defined by applying threshold-based criteria to the TAWSS and OSI fields. Although numerous studies agree that low TAWSS and high OSI are associated with restenosis risk, the thresholds used in the literature vary considerably \cite{hoogendoorn2020multidirectional, Ninno2024modelling_lower_limb, ninno2025silico, de2024predicting, ninno2023systematic}. For this reason, our aim was not to determine the optimal threshold, but rather to demonstrate that our ML framework can robustly predict critical regions under different plausible criteria. To this end, we considered four critical regions defined by two types of thresholding approaches. The first approach applied artery-specific (or relative) thresholds: TAWSS33 for values below the 33rd percentile and OSI66 for values above the 66th percentile, which have been used in multiple studies, including, involving femoral, coronary and other arteries \cite{hoogendoorn2020multidirectional, Ninno2024modelling_lower_limb, ninno2025silico, de2024predicting}. The second approach applied absolute thresholds: TAWSS0.5 (TAWSS $<0.5$ Pa) and OSI0.2 (OSI $>0.2$), commonly adopted in the literature  \cite{ninno2023systematic}. Each threshold produced a binary representation of critical regions, which served as a separate FOM for the subsequent reduced-order and ML analysis.

\appendix
\section*{Appendix A: Derivation of the stenosis deformation function} \label{appendix:A}

\begin{figure}
    \centering
    \includegraphics[width=1\textwidth]{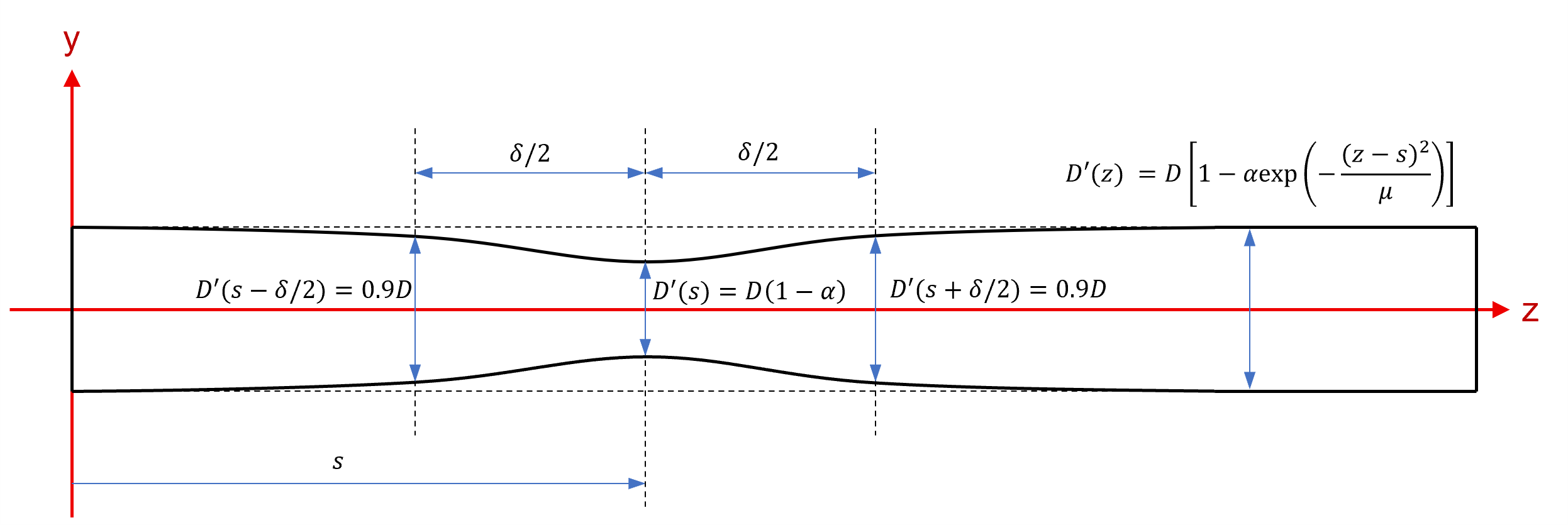}
    \caption{Schematic of Gaussian-like variation in diameter used to introduce stenosis.}
    \label{fig:stenosis}
\end{figure}

The stenosis deformation applied to the baseline mesh was designed to introduce a smooth, Gaussian-like constriction centred at axial location $s$ with severity $\alpha$ and length $\delta$ (Figure \ref{fig:stenosis}). The general form was assumed as

\begin{equation}
    D'(z) = D \left[1 - \alpha \exp\left(-\frac{(z-s)^2}{\mu}\right)\right],
    \label{eq:gaussian_general}
\end{equation}

where $D$ is the local diameter before deformation and $\mu$ is a parameter controlling the width of the Gaussian profile.

\subsection*{Definition of stenosis severity}
At the centre of the stenosis ($z=s$), the diameter reduction is maximal:

\begin{equation}
    D'(s) = D (1 - \alpha).
\end{equation}

Hence, $\alpha$ directly defines the fractional narrowing at the stenosis throat.

\subsection*{Definition of stenosis length}

We know that $\mu$ affects the length of the stenosis $\delta$, but it is difficult to interpret. Our goal here is to rewrite Equation \ref{eq:gaussian_general} in terms of $\alpha$, $\delta$, and $s$. In other words, we would like to eliminate $\mu$ by expressing it in terms of $\alpha$ and $\delta$.

The length of the stenosis, $\delta$, was defined as the full axial distance over which the narrowing decays from $\alpha$ at the throat to $10\%$ at the edges. That is,

\begin{equation}
\begin{aligned}
    \frac{D - D'(s \pm \tfrac{\delta}{2})}{D} &= 0.1 \\
    D'(s \pm \tfrac{\delta}{2}) &= 0.9D
\end{aligned}
\label{eq:define_delta}
\end{equation}

Substituting Equation \ref{eq:define_delta} into Equation \ref{eq:gaussian_general} (with $z = s \pm \delta/2$) gives

\begin{equation}
\begin{aligned}
    0.9D &= D \left[1 - \alpha \exp\left(-\frac{(s \pm \tfrac{\delta}{2}-s)^2}{\mu}\right)\right] \\
    0.9 &= 1 - \alpha \exp\left(-\frac{(\delta/2)^2}{\mu}\right)\\
    0.10 &= \alpha \exp\left(-\frac{\delta^2}{4\mu}\right).
\end{aligned}
\end{equation}

\subsection*{Solving for $\mu$}
Rearranging yields

\begin{equation}
\begin{aligned}
    \exp\left(-\frac{\delta^2}{4\mu}\right) &= \frac{0.10}{\alpha} \\
    -\frac{\delta^2}{4\mu} &= \ln\left(\frac{0.1}{\alpha}\right),
\end{aligned}
\end{equation}

which gives

\begin{equation}
    \mu = -\frac{\delta^2}{4 \ln(0.10/\alpha)}.
\end{equation}

\subsection*{Final form of the stenosis equation}
Substituting $\mu$ back into Equation \eqref{eq:gaussian_general}, the stenosis scaling becomes

\begin{equation}
    D'(z) = D \left(1 - \alpha \exp\left[\left(\frac{z-s}{\delta}\right)^2 \cdot 4 \ln\left(\frac{0.10}{\alpha}\right)\right]\right).
    \label{eq:stenosis_final}
\end{equation}

This is the form implemented in the mesh-deformation algorithm (Eqs.~\ref{eq:stenosis_x}--\ref{eq:stenosis_z} in Section~\ref{sec:Synthetic geometry generation}). The factor $0.10$ specifies that the stenosis length $\delta$ is defined between the two axial points where the narrowing falls to 10\% of its peak value, while the factor 4 arises because $\delta$ spans the full width from $s-\delta/2$ to $s+\delta/2$.

\appendix

\section*{Appendix B: Derivation of the curvature deformation function} \label{appendix:B}

\begin{figure}
    \centering
    \includegraphics[width=0.7\textwidth]{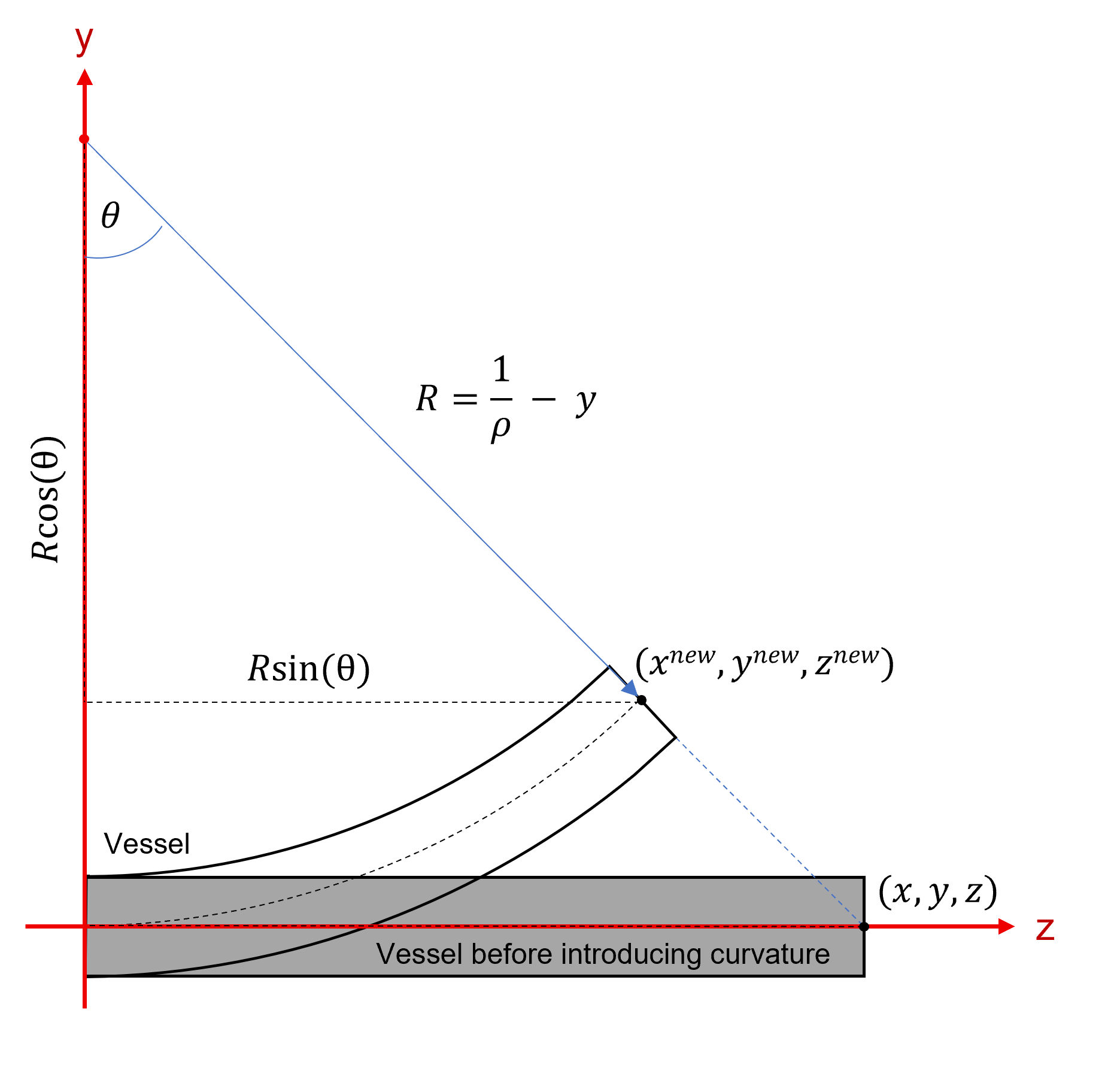}
    \caption{Schematic of dimensions and arc length used to introduce vessel curvature.}
    \label{fig:curvature}
\end{figure}

The curvature deformation maps a straight vessel centreline into a circular arc lying in the $y$–$z$ plane, as shown in Figure \ref{fig:curvature}. The deformation is controlled by the curvature parameter $\rho$ (the inverse of the radius). A node initially located at $(x'', y'', z'')$ is shifted to a new position $(x''', y''', z''')$ that lies on the arc of radius $R$.

\subsection*{Radius of curvature}
The effective local radius of curvature is written as
\begin{equation}
    R = \frac{1}{\rho} - y'',
    \label{eq:appendixB_R}
\end{equation}
such that nodes at larger $y''$ (further from the inner wall) undergo a stronger displacement. This formulation ensures consistent bending across the vessel cross-section.

\subsection*{Angular displacement}
The axial coordinate $z''$ is interpreted as the arc length along the bent centreline. The associated angular displacement is
\begin{equation}
    \theta = \rho z'',
    \label{eq:appendixB_theta}
\end{equation}
with $\rho$ governing the curvature magnitude: higher $\rho$ corresponds to tighter bending. The angle $\theta$ is measured relative to the midline, where the arc length is preserved.

\subsection*{Coordinate transformation}
As in Figure \ref{fig:curvature}, applying the uniform bending yields the following transformation:

\begin{subequations}
\begin{align}
    x''' &= x'', \label{eq:appendixB_x}\\
    y''' &= y'' + R \big(1 - \cos\theta \big), \label{eq:appendixB_y}\\
    z''' &= R \sin\theta. \label{eq:appendixB_z}
\end{align}
\end{subequations}

The $x$ coordinate remains unchanged because the deformation acts solely in the $y$–$z$ plane. The terms $R(1 - \cos\theta)$ and $R\sin\theta$ represent the radial displacement introduced by curvature.

\printbibliography


\maketitle

\section{Hyperparameters tuning results}

\begin{table}[H]
\centering
\caption{Optimal hyperparameters identified by Optuna for the Fourier models across the four critical regions: TAWSS33, OSI66, TAWSS0.5, and OSI0.2.}
\label{tab:hyperparams_Fourier}
\begin{tabular}{lcccc}
\hline
\textbf{Parameter} & \textbf{TAWSS33} & \textbf{OSI66} & \textbf{TAWSS0.5} & \textbf{OSI0.2} \\
\hline
Number of hidden layers         & 4 & 3 & 6 & 5 \\
Number of neurons per layer     & 800 & 1200 & 600 & 600 \\
Number of layers with $L_2$ regularization    & 2 & 2 & 2 & 3 \\
$L_2$ regularization weight   & 1.29e-3 & 1.93e-3 & 1.66e-3 & 1.92e-4 \\
Learning rate       & 1.59e-3 & 7.61e-3 & 2.69e-4 & 3.91e-3 \\
Number of POD modes (output)  & 40 & 135 & 80 & 100 \\
Number of Fourier coefficients (input) & 1 & 3 & 3 & 1 \\
Threshold           & 0.353 & 0.402 & 0.126 & 0.258 \\
\hline
\end{tabular}
\end{table}

\begin{table}[H]
\centering
\caption{Optimal hyperparameters identified by Optuna for the LSTM models across the four critical regions: TAWSS\_low (relative), OSI\_high (relative), TAWSS\_low (absolute), and OSI\_high (absolute).}
\label{tab:hyperparams_LSTM}
\begin{tabular}{lcccc}
\hline
\textbf{Parameter} & \textbf{TAWSS33} & \textbf{OSI66} & \textbf{TAWSS0.5} & \textbf{OSI0.2} \\
\hline
Number of dense layers         & 4 & 2 & 3 & 3 \\
Neurons per dense layer        & 200 & 200 & 600 & 600 \\
Dense layers with $L_2$ regularization   & 3 & 1 & 1 & 2 \\
$L_2$ regularization weight (dense)      & 2.46e-4 & 2.36e-2 & 1.58e-3 & 2.35e-5 \\
Number of LSTM layers          & 2 & 3 & 3 & 4 \\
Neurons per LSTM layer         & 300 & 700 & 500 & 750 \\
LSTM layers with $L_2$ regularization    & 1 & 2 & 2 & 4 \\
$L_2$ regularization weight (LSTM)       & 2.13e-3 & 1.24e-4 & 5.80e-2 & 6.65e-2 \\
Number of combine layers       & 1 & 2 & 2 & 2 \\
Neurons in combine layers      & 800 & 800 & 1200 & 800 \\
Learning rate                  & 9.71e-4 & 3.36e-3 & 1.88e-3 & 2.32e-4 \\
Number of POD modes (output)   & 20 & 50 & 70 & 65 \\   
Threshold                      & 0.413 & 0.447 & 0.082 & 0.369 \\
\hline
\end{tabular}
\end{table}

\begin{table}[H]
\centering
\caption{Optimal hyperparameters identified by Optuna for the CNN models across the four critical regions: TAWSS33, OSI66, TAWSS0.5, and OSI0.2.}
\label{tab:hyperparams_CNN}
\begin{tabular}{lcccc}
\hline
\textbf{Parameter} & \textbf{TAWSS33} & \textbf{OSI66} & \textbf{TAWSS0.5} & \textbf{OSI0.2} \\
\hline
Number of dense layers         & 3 & 1 & 4 & 2 \\
Neurons per dense layer        & 200 & 800 & 600 & 1000 \\
Dense layers with $L_2$ regularization   & 2 & 1 & 0 & 2 \\
$L_2$ regularization weight (dense)      & 5.48e-4 & 7.66e-2 & 4.60e-4 & 2.56e-3 \\
Number of conv. layers         & 4 & 4 & 4 & 4 \\
Number of filters (CNN)        & 400 & 300 & 700 & 300 \\
Kernel size                    & 5 & 11 & 9 & 3 \\
Conv. layers with $L_2$ regularization   & 4 & 4 & 4 & 4 \\
$L_2$ regularization weight (CNN)        & 2.96e-3 & 8.67e-3 & 1.47e-2 & 4.32e-2 \\
Number of combine layers       & 2 & 3 & 2 & 1 \\
Neurons in combine layers      & 1200 & 1200 & 400 & 600 \\
Learning rate                  & 2.45e-3 & 1.44e-3 & 2.98e-3 & 1.39e-2 \\
Number of POD modes (output)   & 55 & 45 & 75 & 105 \\
Threshold                      & 0.365 & 0.403 & 0.097 & 0.173 \\
\hline
\end{tabular}
\end{table}

\section{Training curves}

\begin{figure}[H]
\centering
\includegraphics[width=\textwidth]{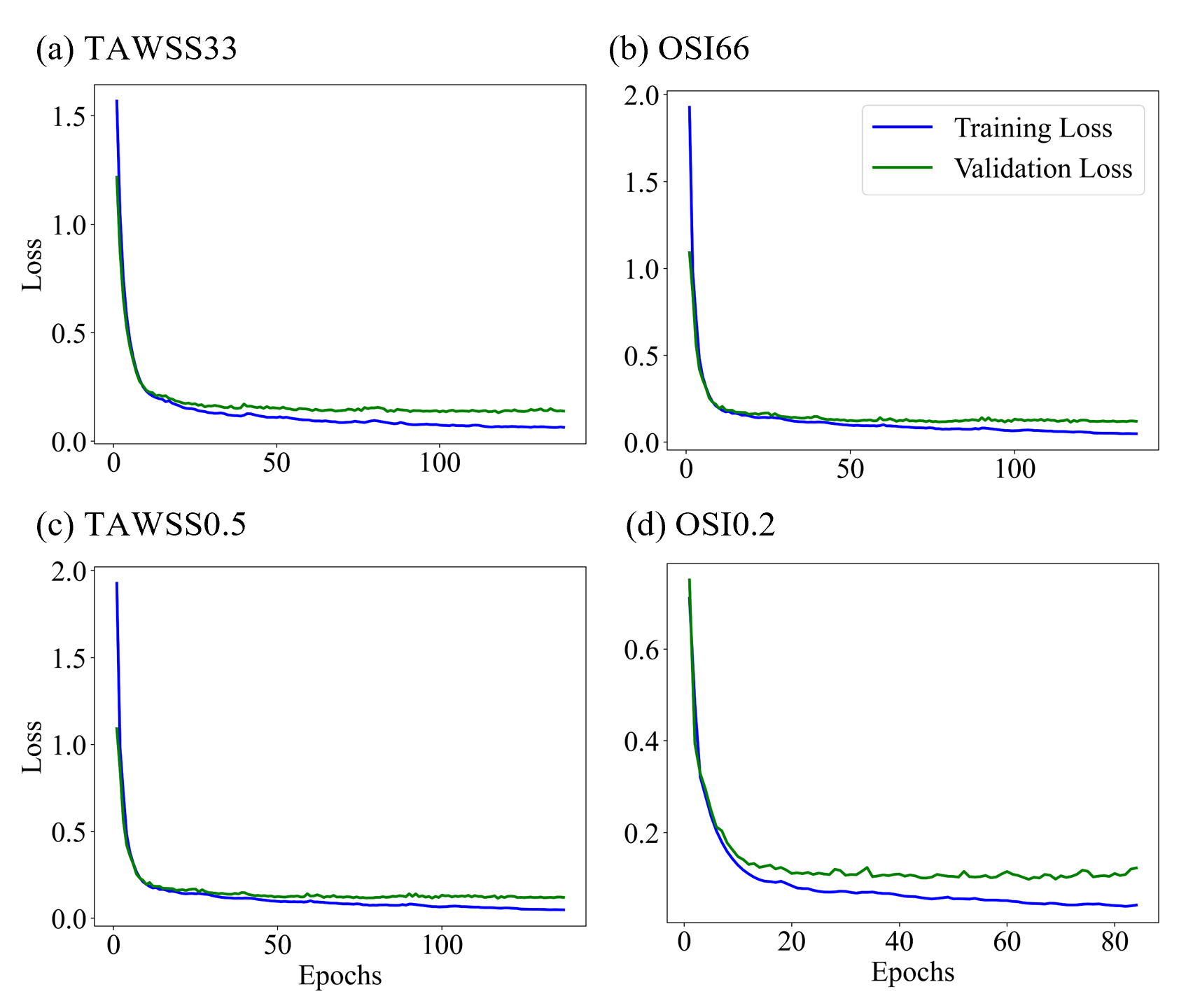}
\caption{
Training and validation loss curves for all three models for the best-performing runs (with the optimal hyperparameters found from Optuna) across all four critical region definitions.}
\label{fig:training_curve}
\end{figure}

\section{BA vs geometric and waveform parameters}

\begin{figure}[ht]
\centering
\includegraphics[width=\textwidth]{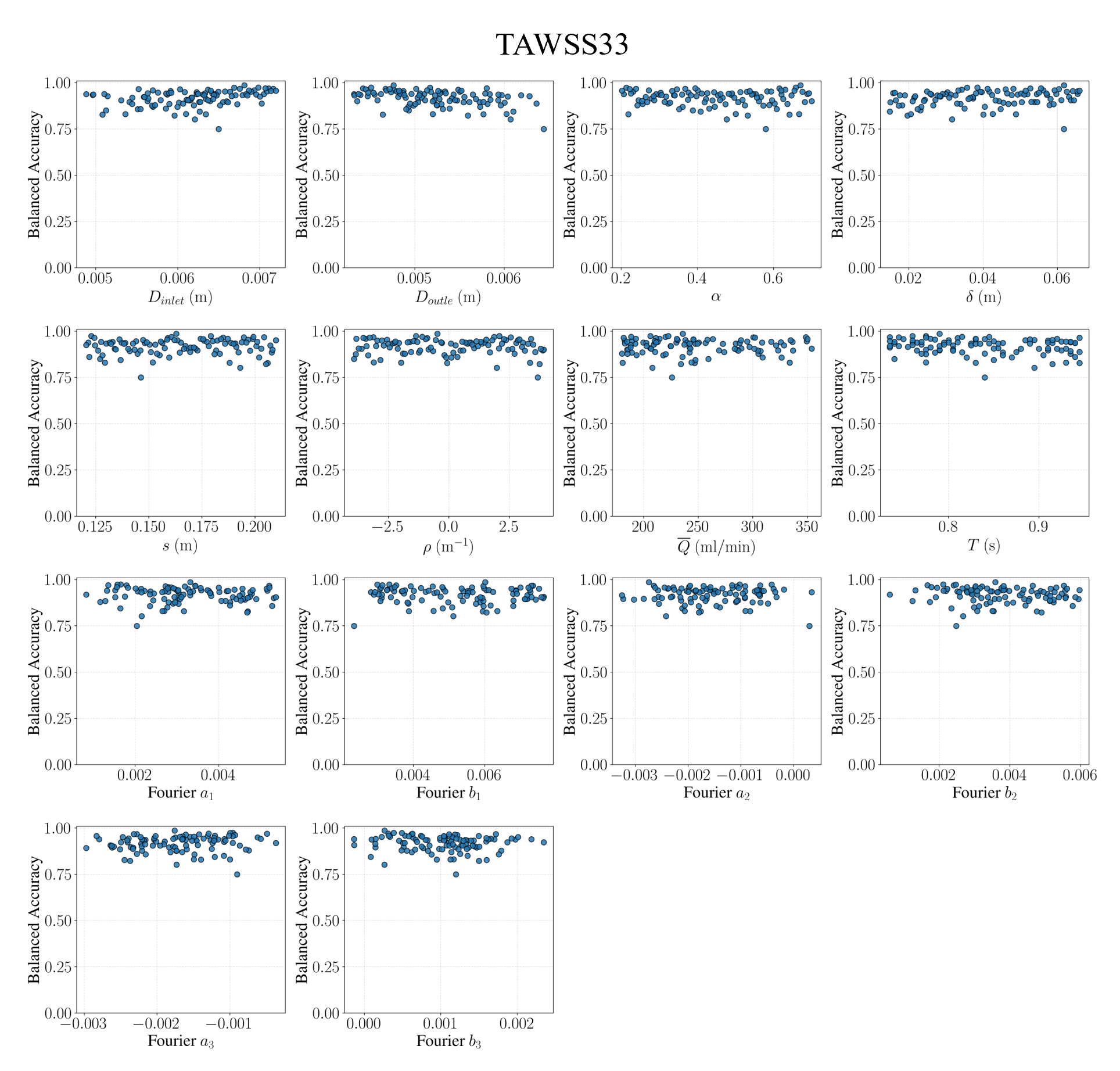}
\caption{
BA of the Fourier models across the test dataset plotted against geometric and waveform parameters for the four critical regions for TAWSS33}
\label{fig:TAWSS33}
\end{figure}

\begin{figure}[ht]
\centering
\includegraphics[width=\textwidth]{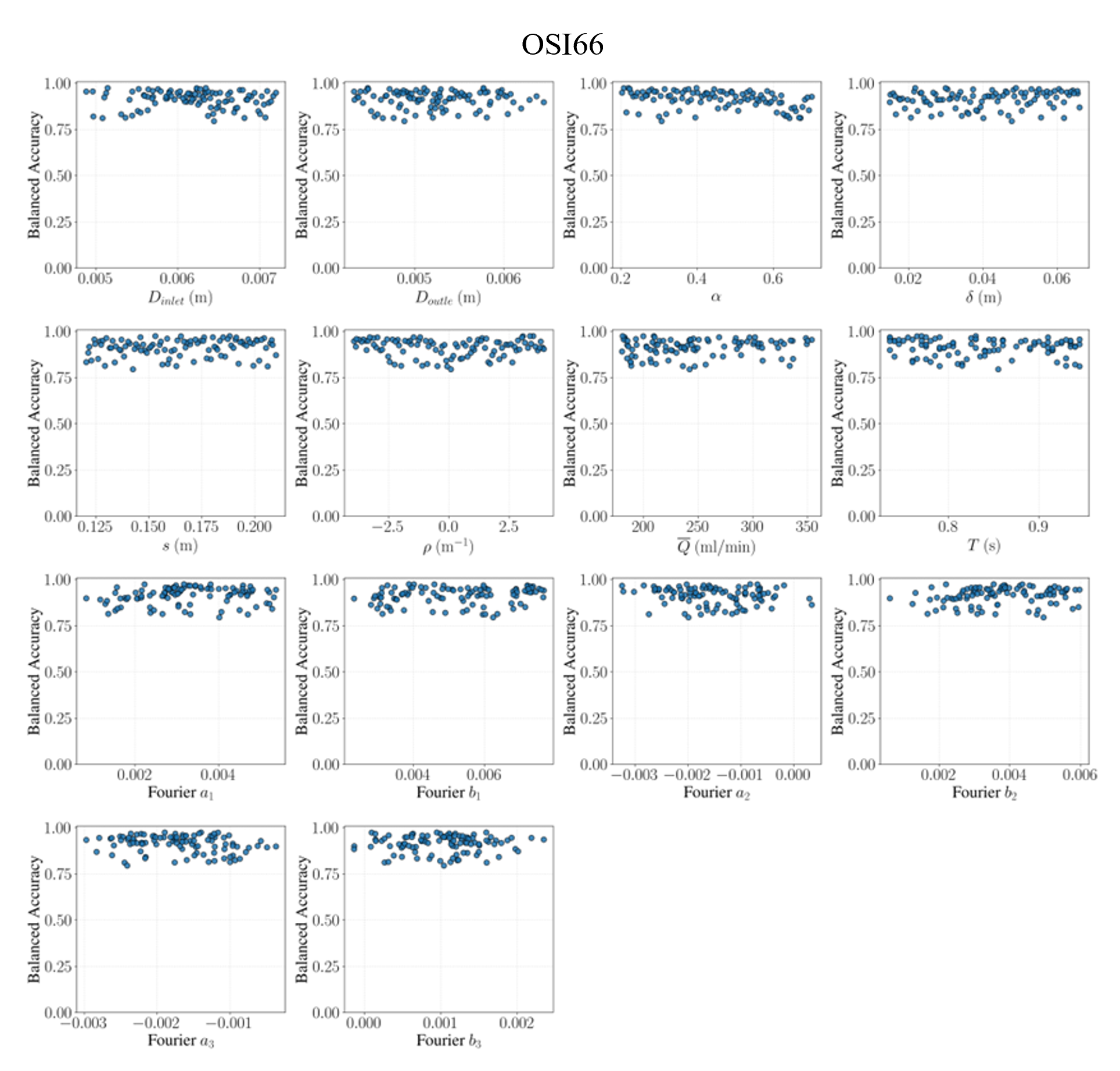}
\caption{
BA of the Fourier models across the test dataset plotted against geometric and waveform parameters for the four critical regions for OSI66}
\label{fig:OSI66}
\end{figure}

\begin{figure}[ht]
\centering
\includegraphics[width=\textwidth]{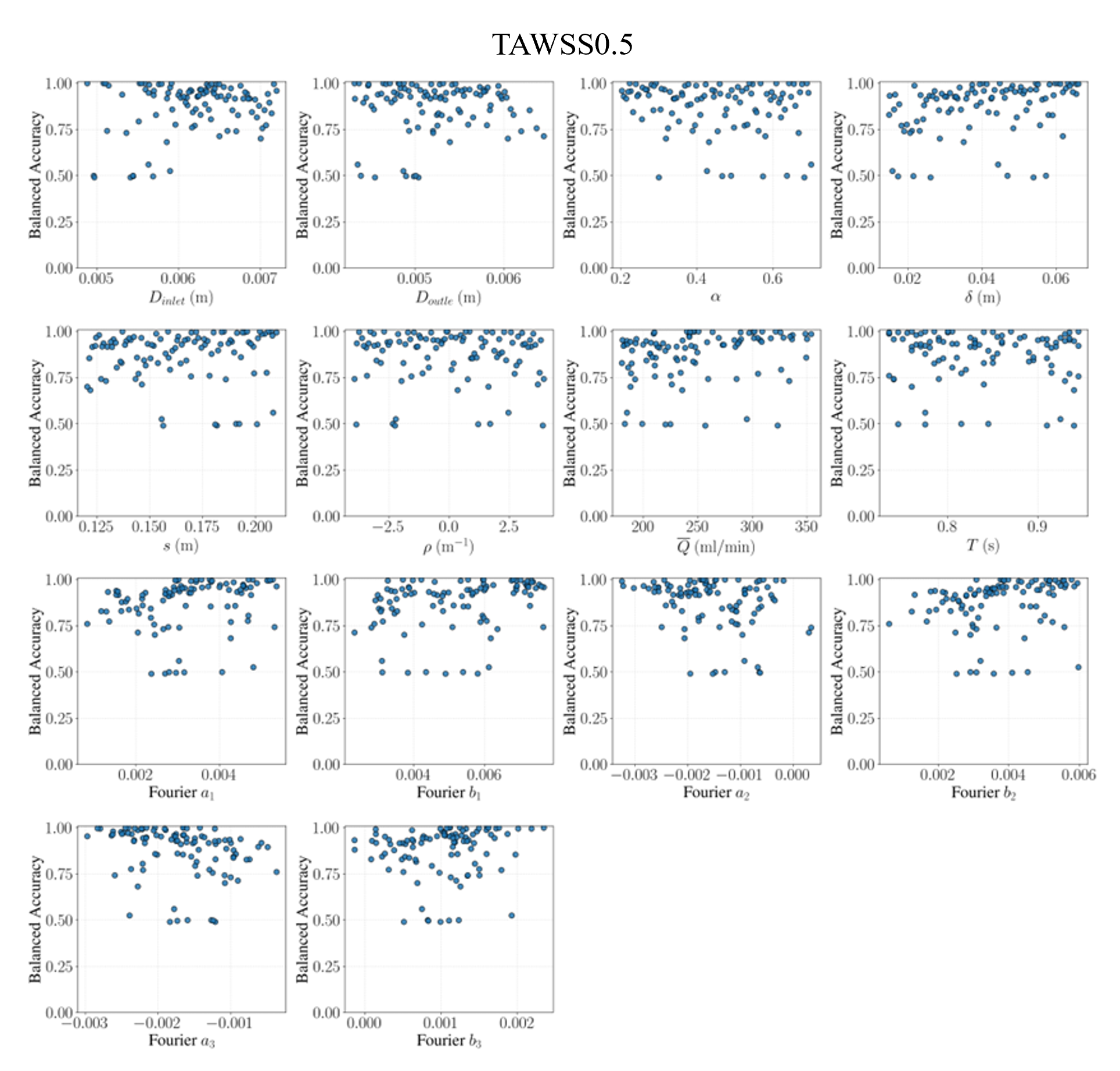}
\caption{
BA of the Fourier models across the test dataset plotted against geometric and waveform parameters for the four critical regions for TAWSS0.5}
\label{fig:TAWSS0}
\end{figure}

\begin{figure}[ht]
\centering
\includegraphics[width=\textwidth]{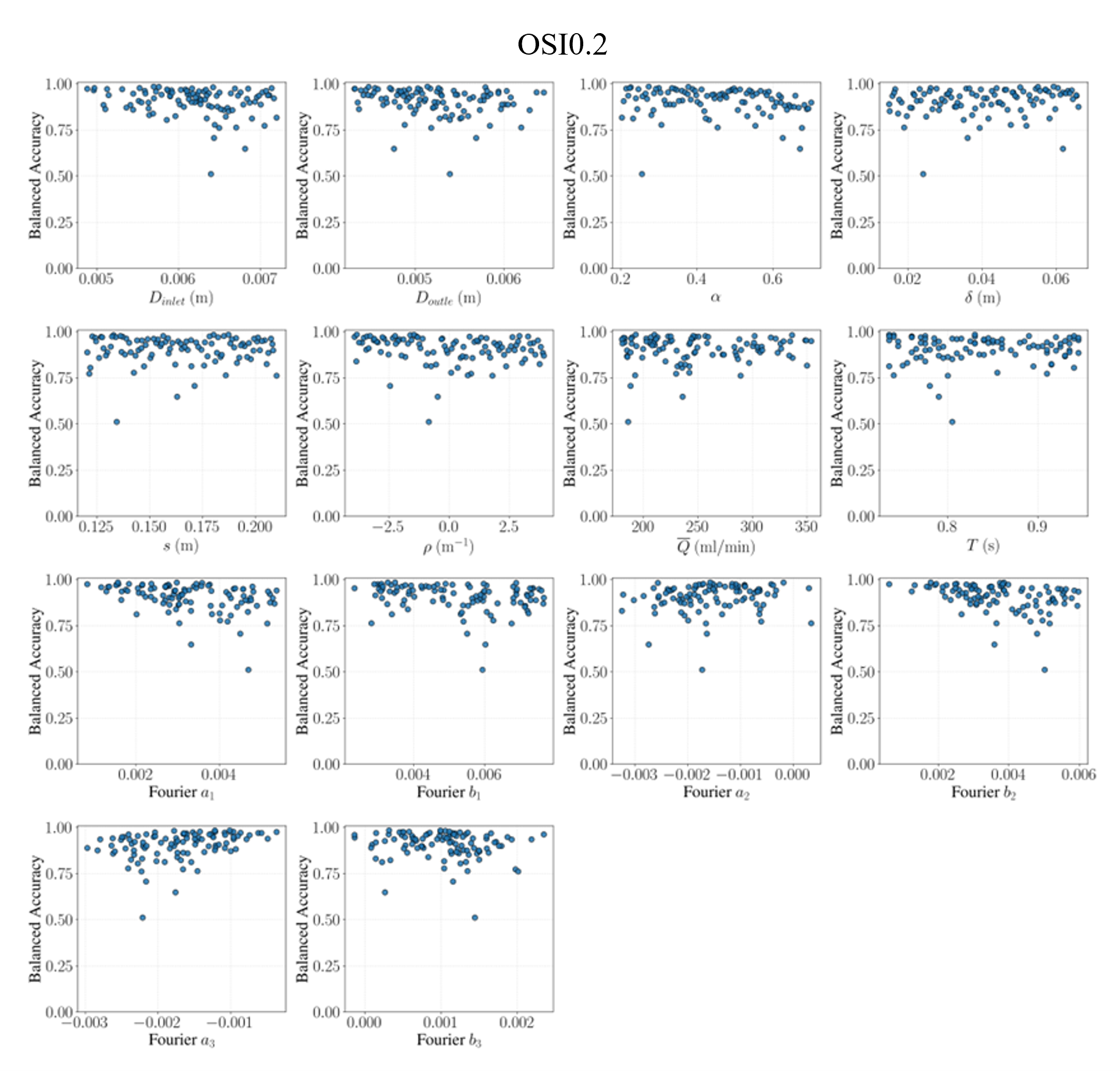}
\caption{
BA of the Fourier models across the test dataset plotted against geometric and waveform parameters for the four critical regions for OSI0.2}
\label{fig:OSI0}
\end{figure}
